\documentclass[10pt,leqno]{amsart}
\usepackage{graphicx}
\usepackage{indentfirst,csquotes}
\usepackage{rotating}
\usepackage{multirow}
\usepackage{enumitem}
\usepackage{subcaption}
\usepackage{graphicx}   
\usepackage{cancel}
\usepackage{soul}
\usepackage{amsmath}
\usepackage{caption}
\usepackage{natbib}
\usepackage{amssymb,amsthm,amsmath}
\usepackage{xcolor,paralist,hyperref,titlesec,fancyhdr,etoolbox}

\titleformat{\section}
  {\normalfont\huge\bfseries}
  {\thesection}
  {1em} 
  {}

\titlespacing*{\section}{0pt}{0ex}{0ex}

\titleformat{\subsection}
  {\normalfont\Large\bfseries\centering}
  {\thesubsection}
  {1em} 
  {\Large}

\titlespacing*{\subsection}{0pt}{1em}{1em} 

\titleformat{\subsubsection}
  {\normalfont\large\bfseries\centering}
  {\thesubsubsection}
  {1em} 
  {\large}

\titlespacing*{\subsubsection}{0pt}{1em}{1em} 

\hypersetup{
  colorlinks=true,
  linkcolor=black,
  filecolor=black,
  urlcolor=black
}

\usepackage{lipsum}

\begin{document}
\title{Dynamic Photometric Variability in Three Young Brown Dwarfs in Taurus: Detection of Optical Flares with TESS data}

\author[Samrat Ghosh]{$^{1,2}$Samrat Ghosh \href{https://orcid.org/0000-0003-3354-850X}{\includegraphics[]{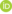}}}

\author[Soumen Mondal]{$^2$Soumen Mondal \href{https://orcid.org/0000-0003-1457-0541}{\includegraphics[]{Orcid.png}}}

\author[Somnath Dutta]{$^3$Somnath Dutta \href{https://orcid.org/0000-0002-2338-4583}{\includegraphics[]{Orcid.png}}}

\author[Rajib Kumbhakar]{$^2$Rajib Kumbhakar \href{https://orcid.org/0000-0001-7277-2577}{\includegraphics[]{Orcid.png}}}

\author[Ramkrishna Das]{$^2$Ramkrishna Das \href{https://orcid.org/0000-0002-5440-7186}{\includegraphics[]{Orcid.png}}}

\author[Santosh Joshi]{$^1$Santosh Joshi \href{https://orcid.org/0009-0007-1545-854X}{\includegraphics[]{Orcid.png}}}

\author[Sneh Lata]{$^1$Sneh Lata \href{https://orcid.org/0000-0001-9367-1580}{\includegraphics[]{Orcid.png}}}
\maketitle
\begin{center}

$^1$Aryabhatta Research Institute of Observational Sciences (ARIES), Manora Peak, Nainital-263002, India\\
$^2$S N Bose National Centre for Basic Sciences, Salt Lake, Kolkata 700 106, India\\ 
$^3$Institute of Astronomy and Astrophysics, Academia Sinica, Taipei 10617, Taiwan\\

\end{center}

\let\thefootnote\relax

\begin{abstract}
We present $I$-band time-series photometric variability studies of three known nearby ($\sim$ 140 pc) and young ( $\sim$ 1 Myr) brown dwarfs (BD) in the Taurus star-forming region in the Perseus Molecular Cloud. From 10 nights of observations over a time span of 10 years, with a typical run of 3 to 6 hours each night, we estimated that the BDs show unstable short-scale periodicity from 1.5 to 4.8 hours. Using the long-term photometry from the Transiting Exoplanet Survey Satellite (TESS), we have conducted a time-resolved variability analysis of CFHT-BD-Tau 3 and CFHT-BD-Tau 4, revealing orbital periods of $\sim$ 0.96 days and $\sim$ 3 days respectively, consistent with earlier studies. We also found two superflares in TESS sector 43 data for CFHT-BD-Tau 4 and estimated the flare energies as $7.09\times10^{35}$ erg and $3.75\times10^{36}$ erg. A magnetic field of $\sim3.39 ~kG$ is required to generate such flare energies on this BD. We performed spot modelling analysis on CFHT-BD-Tau 3 and CFHT-BD-Tau 4 to address the variability detected in the data using the package BASSMAN. Spectral energy distribution and infrared colours of the sources suggest that they have a sufficient amount of circumstellar material around them.
\end{abstract}



\section{Introduction}

Brown Dwarfs (BDs) are sub-stellar objects with a mass ranging from 13 $M_J$ ($M_J:$ Jupiter Mass) to 75 $M_J$  for solar metallicity. Choosing a star-forming region seems a plausible choice for studying these objects as their luminosities and temperatures are higher than their counterparts in the galactic field. Taurus is a well-characterized star-forming region for the surveys of BDs and very low-mass stars (VLMs) for its proximity ($\sim 140 ~pc$; \citealt{1994AJ....108.1872K};  \citealt{1998MNRAS.301L..39W} ) and its rich population of young stellar objects (YSOs) (\citealt{2008hsf1.book..405K}; \citealt{2016yCat..21860111L}, \citealt{2017AJ....153...46L}) detected with optical and infrared (IR) monitoring. Stars are forming in non-clustered isolated systems (density of stars, n $\sim 1-10 ~pc^{-3}$) relative to denser star-forming regions (SFRs) like Orion Nebula Cluster (ONC), and relatively free from massive stars nearby \citep{Briceno2008}. It is nearer than other regions, so fainter and lower mass members can be detected with 1-2 m class telescopes and for low extinction ($\rm A_v \sim 4$ mag) of most of the members of Taurus \citep{Briceno_2002}, optical photometric observations are suitable in this region.

\begin{figure}
\includegraphics[height=6in]{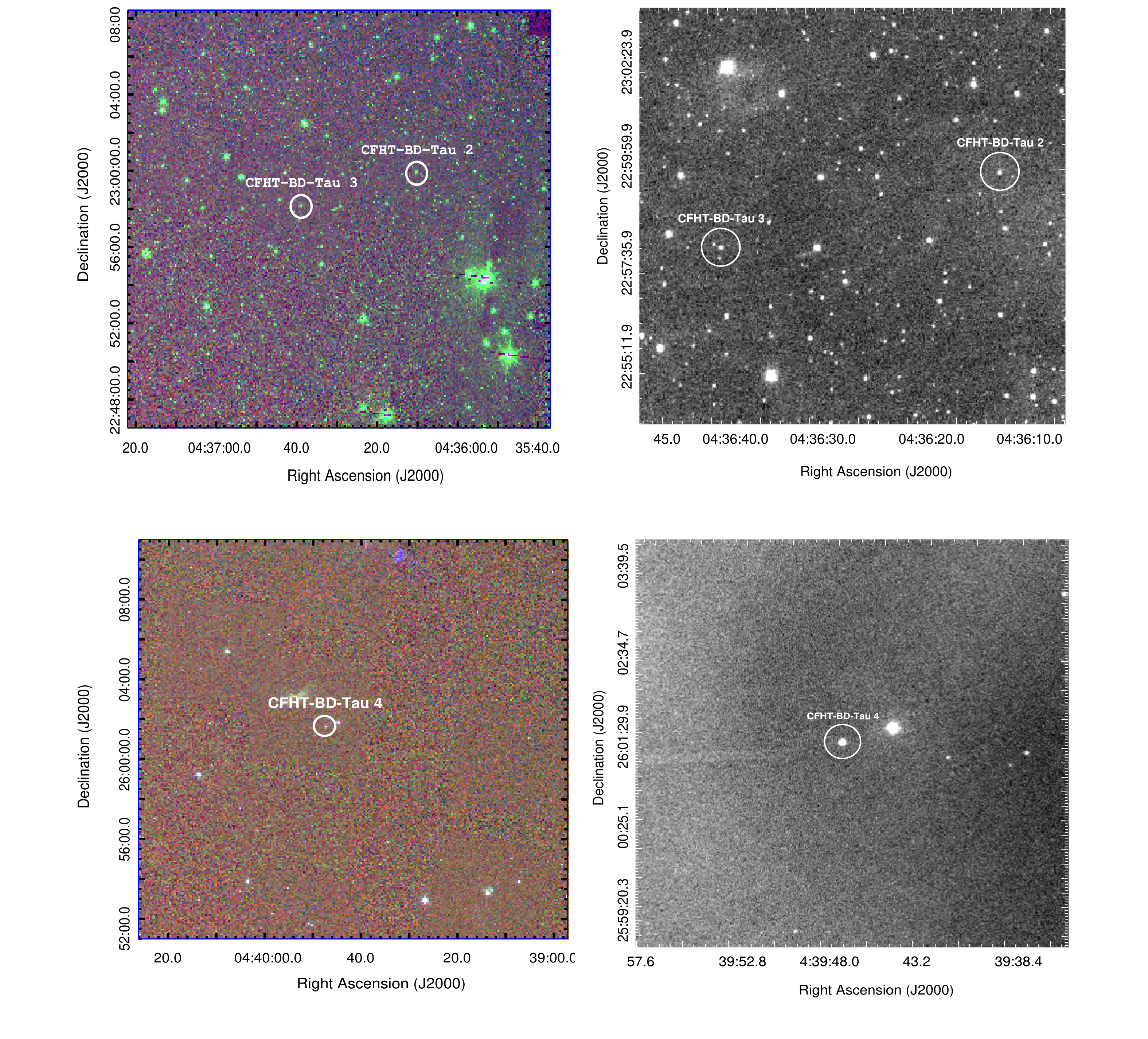}
\caption{{\bf Top Left}: Spitzer-IRAC colour composite image of the $\rm 20\times20 ~arcmin^2$ region showing CFHT-BD-Tau 2 and CFHT-BD-Tau 3 generated with  IRAC 3.6 $\mu m$ (blue), IRAC 4.5 $\mu m$ (green), and IRAC 8.0 $\mu m$ (red). The studied objects are marked with white circles. {\bf Top Right}:  An image of $I$-band observations showing CFHT-BD-Tau 2 and CFHT-BD-Tau 3 on 8th December 2018 observed with 1.3-m DFOT operated by ARIES, Nainital. {\bf Bottom Left:} Spitzer-IRAC colour composite image of the $\rm 20\times20 ~arcmin^2$ region showing CFHT-BD-Tau 4 marked with a white circle, generated with IRAC 3.6 $\mu m$ (blue), IRAC $4.5 \mu m$ (green), and IRAC 8.0 $\mu m$ (red). $\rm 10 \times 10 arcmin^2$. {\bf Bottom Right:}  Image of $I$-band observations showing CFHT-BD-Tau 4 is shown here. Data is taken in 1.04-m ST operated by ARIES, Nainital on 23rd October 2009.}
\label{region}
\end{figure}

At a young age (a few Myr), BDs and VLMs are detectable via X-rays, UV, mid-IR, and optical wavelengths. Photometric variability studies can detect periodic and aperiodic variability in young BDs and probe the nature of their atmospheres \citep{2014ApJ...796..129C, Wilson2014, Marley2015}. Variability in brown dwarfs (BDs) traces the physical condition of their circumstellar environment. Stable surface features like starspots or dust clouds are the reason for non-uniformity in its surface, which causes periodic optical modulation of the flux as the object rotates. Whereas aperiodic or irregular variability might be due to non-uniform accretion from circumstellar disks, magnetic spots, dynamic star spots and dust clouds, changing with different time scales than their rotation periods. 

BDs are fast rotators with periods ranging from a few hours to days \citep{2010ApJS..191..389C, 2014yCat..35660130C}. The average rotation rate of stars varies across the main sequence, differing in stellar structure and magnetic properties. Main-sequence stars initially spin down by losing angular momentum through magnetic braking \citep{1967ApJ...148..217W}. At both ends of the main sequence, sharp increases in the rotation rate (\citealt{10.2307...40678841}) are observed. At the lower end of the main sequence, late-type M dwarfs, including BDs, rotate rapidly for being fully convective. For lacking the radiative core, these objects are thought to produce topologically different magnetic fields, which may not be ready to efficiently dissipate momentum \citep{2013EAS....62..143B, 2017ApJ...837...96H}. The periodic variability can be observed within a few nights of photometric monitoring using 1-2 m class ground-based telescopes. 

We have monitored 3 known brown dwarfs: CFHT-BD-Tau 2, CFHT-BD-Tau 3 and CFHT-BD-Tau 4 (Age $\sim$ 1 Myr \citep{2001ApJ...561L.195M}; hereafter CT2, CT3 and CT4, respectively) in Taurus Molecular clouds with different Indian telescope facilities on various epochs. BD nature of CT2, CT3 and CT4 has been confirmed spectroscopically \citep{2001ApJ...561L.195M}. We have also conducted a time-resolved variability analysis of CT4 using the 2-min cadence data from the Transiting Exoplanet Survey Satellite (TESS). TESS observed CT4 in September 2021 and October 2021 (TIC 150058662; TIC: TESS Input Catalogue ID; \citealt{2021arXiv210804778P}) in sectors 43 and 44 and CT3 in September 2023 and October 2023 (TIC 118555907) in sector 70 and 71. TESS has observed CT2 with TIC 118554419 in sectors 43 and 44, though these observations are part of the TESS collateral pixel data, which do not have any illumination on them. Therefore they do not collect any data on the targets. These columns are used for calibration in the TESS science pipeline. CT4 has been observed in X-rays, mid-IR to sub-mm wavelengths \citep{2003ApJ...585..372L, 2003ApJ...590L.111P, 2004A&A...426L..53A}, and confirmed to have a circumstellar disk which has been modelled extensively and the sub-mm wavelength data establishes the disk mass 0.4–2.4 $M_J$ (\citealt{2003ApJ...593L..57K}). Detection of this amount of disk material makes a possibility of planet-formation around BDs and gives us insight into BD's formation \citep{2006ApJ...645.1498S}. Previous studies \citep{2004A&A...426L..53A} revealed that the CT4 disk is dominated by an order of magnitude larger 2.0 $\mu m$ grain than commonly found in interstellar matter. Dust settling is also taking place in the disk of CT4. These two processes are the necessary steps for forming planets around stars (\citealt{1993ARA&A..31..129L}). CT4 appears to be relatively brighter than similar 1 Myr isochrones objects, whereas CT2 and CT3 are intrinsically fainter than CT4 \citep{2001ApJ...561L.195M} and had no confirmed accretion. Previous studies revealed that CT2, CT3 and CT4 have day-scale rotation periods of 2.93, 0.96 and 2.95 days, respectively \citep{Scholz_2018, 2020AJ....159..273R} in K2 mission observations.

CT4 has shown very energetic flare emissions in previous studies with energies up to $2.1 \times 10^{38}$ erg from Kepler K2 optical data (\citealt{2018ApJ...861...76P}), which suggests chromospheric activity. Activity in CT4 could be inferred from strong $H_\alpha$ with $Br_\gamma$  \citep{2001ApJ...561L.195M} and quiescent X-ray detection \citep{2007MmSAI..78..368G} in previous studies. These two indicators suggest that CT4 inhibits an active chromosphere and corona. Generally, M dwarfs are known to emit high energy flares \citep{2014ApJ...797..122D}, analogous to the sun but more powerful. Flares are multi-wavelength high-energy outbursts, with energy ranging from $E \sim 10^{26} - 10^{38} ~erg$ \citep{Kowalski_2010, Davenport_2016, Schmidt_2019},  while typical solar flare energy is approximately $\sim 5 \times 10^{32}$ erg  \citep{2013JSWSC...3A..31C}. Earlier studies reported that these three BDs have rotation periods ranging $\sim$1 - 3 days \citep{2018ApJ...861...76P, Scholz_2018, 2020AJ....159..273R}. We have provided the details about the sources in Table \ref{source_details}.

\begin{table*}
    \centering
    \caption{Properties of the three studied sources}
    \label{source_details}
\begin{tabular}{ p{3cm} p{2.8cm} p{2.8cm} p{2.8cm} p{4.7cm}}
    \hline
    Parameter & Value & Value & Value & References \\
    \hline
    Object & CFHT-BD-Tau 2 & CFHT-BD-Tau 3 & CFHT-BD-Tau 4 & \\
    Identifier & 2MASS J04361038+2259560 & 2MASS J04363893+2258119 & 2MASS J04394748+2601407 & 2MASS Catalog: \cite{2003tmc..book.....C} \\
    RA (hr:min:secs) & 04:36:10.4 & 04:36:38.9 & 04:39:47.3 & \\
    Dec (deg:min:sec) & +22:59:56 & +22:58:12 & +26:01:39 & \\
    SpT & M7.5 & M7.75 & M7 & \cite{2010ApJS..186..111L} \\
    I (mag) & 16.69 & 16.79 & 15.64 & \cite{2007AA...465..855G} \\
    R (mag) & 19.17 & 19.21 & 19.11 & NOMAD Catalog: \cite{2004AAS...205.4815Z} \\
    J (mag) & 13.754 $\pm$ 0.025 & 13.724 $\pm$ 0.026 & 12.168 $\pm$ 0.023 & 2MASS Catalog: \cite{2003tmc..book.....C} \\
    H (mag) & 12.762 $\pm$ 0.022 & 12.861 $\pm$ 0.024 & 11.008 $\pm$ 0.021 & '' \\
    K (mag) & 12.169 $\pm$ 0.019 & 12.367 $\pm$ 0.025 & 10.332 $\pm$ 0.018 & '' \\
    IRAC [3.6] $\mu m$ (mag) & 11.62 $\pm$ 0.05 & 11.71 $\pm$ 0.05 & 9.38 $\pm$ 0.05 & \cite{2006ApJ...647.1180L} \\
    IRAC [4.5] $\mu m$ (mag) & 11.38 $\pm$ 0.05 & 11.62 $\pm$ 0.05 & 8.97 $\pm$ 0.06 & '' \\
    IRAC [5.8] $\mu m$ (mag) & 11.35 $\pm$ 0.04 & 11.54 $\pm$ 0.04 & 8.54 $\pm$ 0.02 & '' \\
    IRAC [8.0] $\mu m$ (mag) & 11.29 $\pm$ 0.04 & 11.56 $\pm$ 0.04 & 7.79 $\pm$ 0.03 & '' \\
    MIPS [24] $\mu m$ (mag) & 10.61 $\pm$ 0.04 & 8.55 $\pm$ 0.04 & 4.96 $\pm$ 0.04 & \cite{2010ApJS..186..259R} \\
    Period (d) & 2.93 & 0.96 & 2.95 & \citealp{Scholz_2018, 2020AJ....159..273R} \\
    Mass (M$_\odot$) [error$^a$] & 0.044 [50\%]& 0.038 [50\%] & 0.064 [50\%] & \cite{2011ApJ...727...64K, 2018ApJ...861...76P} \\
    Dust Masses ($M_E)^b$ & $<$4.1 & $-$ & 1.4–7.6 & \cite{2003ApJ...593L..57K} \\
    $v \sin i$ [km/s] & 8 & 12 & 11 & \cite{2005ApJ...626..498M} \\ \hline
\end{tabular}

    $^a$Assumption of a 1 Myr constant age and the stellar models are chosen, gives uncertainties in the masses of order 50\% \citep{2011ApJ...727...64K}; $^bM_E \sim$ Mass of the Earth; .
\end{table*}

The paper is organized as follows: section \ref{obs_sect} describes the observations for this study, and section \ref{data_sect} briefly describes the data reduction process. In section \ref{result_sect}, we show the results of our work and discuss the results in section \ref{discussion_sect}. We have summarized our work in section \ref{summary}.

\section{Observations}
\label{obs_sect}
\subsection{Ground-based Observations}

The $I$-band optical photometric data were obtained from various Indian telescope facilities over the years. Data were acquired using 1-m Sampurnanda Telescope (hereafter, 1-m ST) and 1.3-m Devasthal Fast Optical Telescope (hereafter, 1.3-m DFOT) operated by ARIES, Nainital. Data were also obtained from 1.3-m J. C. Bhattacharya Telescope (hereafter, 1.3-m JCBT), Kavalur, Tamilnadu and 2-m Himalayan Chandra Telescope (hereafter, 2-m HCT), Ladakh, operated by the Indian Institute of Astrophysics (IIA), Bangalore. The Wright 2k CCD with 24$~\mu m$ pixel size is used in 1-m ST for acquiring the data with a field of view (FoV) of 13$\times$13 $arcmin^2$ (\citealt{2014PINSA..80..759S}).
The ANDOR $2K\times2K$ CCD instrument is used for our optical observations in 1.3-m DFOT. This instrument has a pixel size of $13.5~\mu m$. A set of Johnson-Cousin $B$, $V$, $R$, $I$, and $H_\alpha$ circular filters are available which gives an unvignetted field of view of $18\times18 ~arcmin^2$. The camera was operated at 1 MHz readout mode with RMS noise of 6.5 $e^{-1}$ and a gain of 2.0 $e^{-1} /ADU$. 1.3-m JCBT is equipped with the $2K\times4K$ CCD having a FoV of $9\times18 ~arcmin^2$ with plate scale 20 arcsec/mm. The backend instrument used in 2-m HCT is the Himalayan Faint Object Spectrograph and Camera (HFOSC). The $2K \times 2K$ part of the detector in $2K \times 4K$ CCD having a pixel size of 15 $\mu m$ and a pixel scale of 0.296 arcsec is used for imaging observations. The FoV on the $2K \times 2K$ part of CCD in the imaging mode is $10 \times 10$ $arcmin^2$. In all of these instruments, photometric images were taken in the optical $I$-band with exposure times varying according to the night condition and altitude of the telescopes. In a few nights, snapshot $I$-band images of CT4 were taken before and after the continuous monitoring of CT2 and CT3. We did this to combine these CT4 data points with other nights for a longer baseline and detect any longer period if it exists. The detailed information about observations is mentioned in Table \ref{log-table}.

\subsection{ TESS: Transiting Exoplanet Survey Satellite}
TESS\footnote{\url{https://archive.stsci.edu/missions-and-data/tess}} is a NASA mission led by MIT to discover transiting exoplanets by sky survey. It operates in the red-optical band-pass, covering the wavelength range from 600 to 1000 nm. The satellite is equipped with wide back-illuminated CCDs, with $4096\times4096$ pixels fitted within $62\times62$ $mm^2$ area. The imaging area is $2048\times2048$ pixels. The remaining pixels are used as a frame-store to allow rapid, shutterless readout ($\sim$4 ms) with readout noise less than 10 $e^-/s$. The CCDs operate at a temperature of around $-75~ ^\circ C$, making the dark current negligible. The data from the TESS are returned to Earth when the spacecraft reaches perigee every 13.7 days\footnote{\url{https://tess.mit.edu/science/}}. High cadence is important for the detection of exoplanets. TESS conducts high-resolution photometry of stars with a cadence of approximately 2 minutes. These data are made available. In addition, the FFI (full-frame image) is read out approximately every 30 minutes. Photometry can be performed on FFIs of any target within the $24\times96$ $degree^2$ FoV.

TESS observed CFHT-BD-Tau 4 with camera 2 during sector 43 from September 16, 2021, to October 11, 2021, and with camera 3 during sector 44 from October 12, 2021, to November 5, 2021. CFHT-BD-Tau 3 was with camera 4 during sector 70 from September 20, 2023, to October 16, 2023, and with camera 4 during sector 71 from October 17, 2021, to November 11, 2023. The four cameras of TESS as backend instruments cover the field of view of each $24\times24 ~deg^2$ and are aligned to cover $24 \times 96 ~deg^2$ of the sky, which is called `sectors’ \citep{2015JATIS...1a4003R}. The data were stored under the Mikulski Archive for Space Telescopes with identification numbers `TIC 150058662’ and `TIC 118555907' for CT4 and CT3. We have considered only the data with the `quality flag' (Q) = 0, as quality flags greater than 0 denote anomalous events in the FITS file data structure. We retrieved TESS data with the TIC of the objects CT3 and CT4 from the Mikulski Archive for Space Telescopes and processed the light curves from TESS 2-min cadence data using ‘lightkurve’ packages \citep{2018ascl.soft12013L}. The 2-min cadence data were processed through the Science Processing Operations Center (SPOC) \citep{2016SPIE.9913E..3EJ} pipeline. We used Pre-Search Data Conditioning (PDCSAP) light curves because these are already corrected for systematic instrumental noise present in the Simple Aperture Photometry (SAP) light curves. The 2-min cadence PDCSAP data are more sensitive to short-timescale flux variation and have less scatter \citep{Smith_2012, Stumpe_2014}.

\renewcommand{\tabcolsep}{5pt} 
\begin{table*}

\fontsize{8}{7}\selectfont
\caption{Observation Log for the three BDs in Taurus}
\label{log-table}
\hspace{-2.2cm}
\begin{tabular}{p{1.4cm}p{2.4cm}p{1.6cm}p{1.8cm}p{1.3cm}p{1.8cm}p{3cm}p{1.2cm}p{1.2cm}}
\hline\\
Date        & Target		& Telescope   & Instrument      	&   FoV ($arcmin^2$) & Run-Length$^a$   (hours) & N$\times$Exposure (s) &  Seeing (arcsec) & Night condition\\ \hline\\
30.01.2009  &	CT4			&	1-m ST 		&  2k Wright ccd	& 13$\times$13 	& 4.10  & 46$\times$300 &  1.14 & dark\\\\
31.01.2009 	&	CT2, CT3			&	'' 			& ''				&  ''   			& 5.81  & 1$\times$400, 17$\times$500 & 1.22  & dark \\\\
23.10.2009	&	CT4			&	''  			& '' 		 	&  ''   			& 3.19 & 32$\times$300 & 1.03 & dark \\\\
08.12.2018 	&	CT2, CT3			&	1.3-m DFOT 	& ANDOR 2Kx2K 	&  18$\times$18 	& 5.47 & 3$\times$300, 39$\times$400 &  3.06 & dark \\\\
09.12.2018  &	CT4			&	 ''			& ''			 	&  ''   			& 5.58 & 27$\times$300  & 2.64 & dark \\\\
01.01.2019  &	CT2, CT3; CT4$^{b}$		&	1.3-m JCBT	& 2kx4k ccd  	& 9$\times$18  	& 4.47; 6.09 & 1$\times$300,1$\times$600, 14$\times$720,41$\times$900; 1$\times$300, 8$\times$360 & 2.1  & dark \\\\
02.01.2019  &	CT2, CT3; CT4$^{b}$		&	''			& 	''			& ''				& 4.72; 0.66 & 1$\times$600, 21$\times$720 ; 1$\times$300, 5$\times$420 & 1.58 & dark \\\\
03.01.2019  & 	CT4$^{**}	$		&	''		 	&    ''  		& ''    			& 5.39  & 10$\times$420 & 1.84 & dark  \\\\
25.12.2019  &   CT2, CT3; CT4$^{b}$     &   2-m HCT      &   2kx2k CCD   & 10$\times$10      &   5.97; 8.42 & 40$\times$200, 58$\times$180; 1$\times$60, 20$\times$90, 21$\times$120 & 1.61 & dark \\\\ 
26.12.2019  &   CT2, CT3; CT4$^{b}$ &  ''     &  ''   & 10$\times$10  &5.00; 6.00   &1$\times$200, 85$\times$180 ; 1$\times$120, 25$\times$100 & 1.61  & dark \\\\  \hline

\end{tabular}

$^a$Total duration of observation. $^{b}$Snapshot observation of CT4 

\end{table*}

\section{Data Analysis}
\label{data_sect}

The raw CCD images from ground-based $I$-band observations were pre-processed using IRAF\footnote{Image Reduction and Analysis Facility (IRAF) is distributed by National Optical Astronomy Observatories (NOAO), USA (http://iraf.noao.edu/)} standard packages. For bias subtraction, flat fielding and cosmic ray removal, standard IRAF packages- {\sc{flatcombine}}, {\sc{ccdprocess}}, {\sc{crutil}} and {\sc{ccdprocess}} were used. All the sources including the BDs were marked in the observed frames using {\sc{IRAF}}'s {\sc{DAOFIND}} \citep{1992JRASC..86...71S} task. A few sources were marked manually using {\sc tvmark} interactively whenever the {\sc{DAOFIND}} task failed to mark due to the faintness of the sources.

\subsection{Aperture Photometry}

Aperture photometry has been performed on the selected target sources as well as other unsaturated sources present in the frame by selecting radii from 1 to 25 pixels using the {\sc phot} \citep{1987PASP...99..191S} task of {\sc iraf}. We selected an aperture from this range (1 to 25 pixels) using the growth curve (instrumental magnitude vs. aperture plot) after which changing the aperture does not change the magnitude of the sources significantly \citep{2021MNRAS.500.5106G}. For different nights, using individual aperture values varying from 8 to 12 pixels depending on the telescopes and the seeing (Table \ref{log-table}) of the observing nights, and using a sky annulus radius of 12 pixels and a width of 5 pixels for background subtraction, we obtained the instrumental magnitude for each source by running the {\sc phot} task for the second time.

\section{Results}
\label{result_sect}

\subsection{Differential Photometry}
\begin{figure*}
\hspace*{-2.0cm}
   \includegraphics[height=7.in]{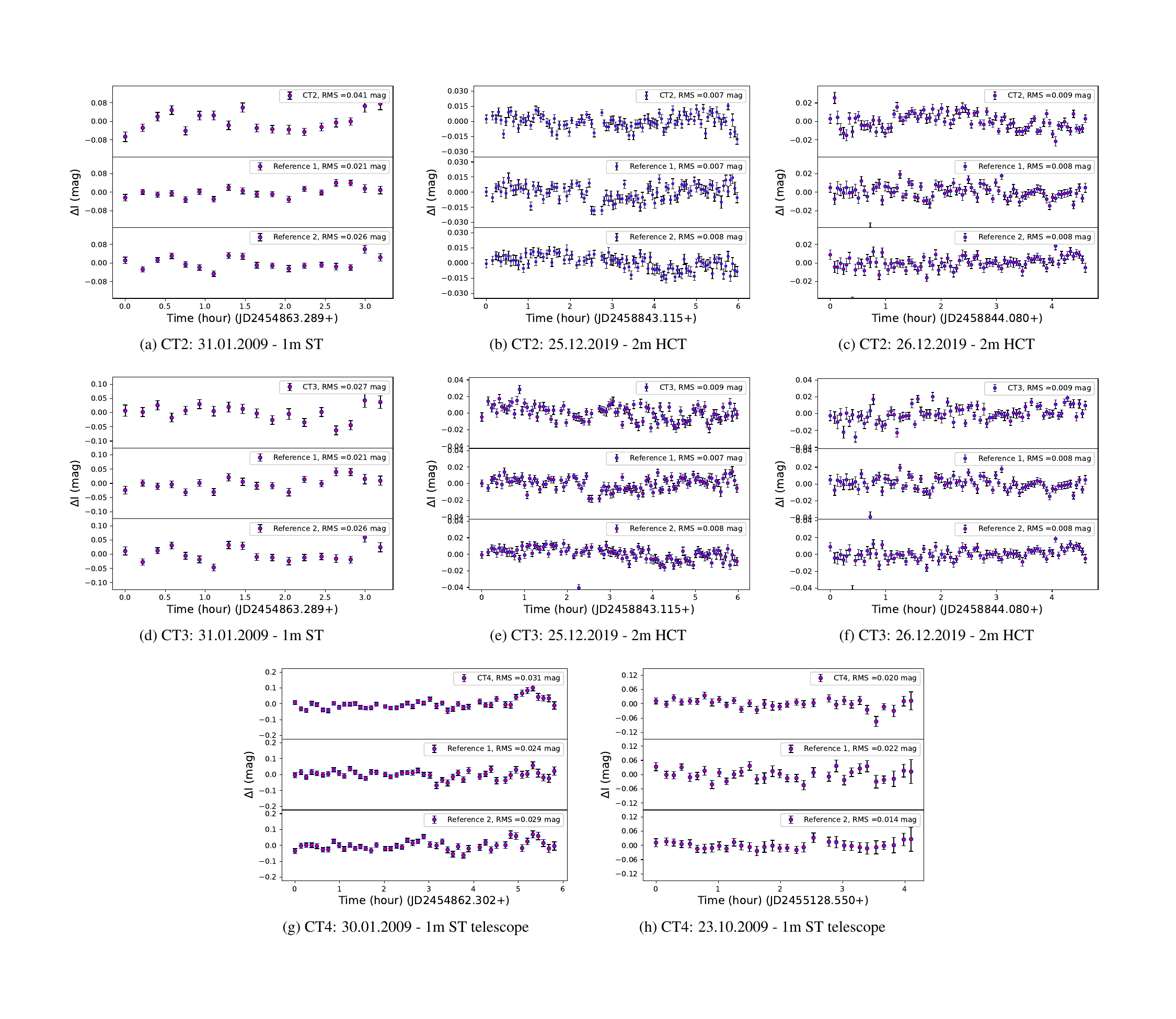}
  \caption{Variable light curves of the three brown dwarfs are shown. Two non-variable reference light curves are shown in the bottom panels of each light curve. The source name, dates and telescopes used to obtain the data, are mentioned underneath each panel.}
\label{lc_plot}
\end{figure*}

For each observing date, we used the estimated instrumental magnitudes to construct the time-series light curves for all sources. Following \cite{2021MNRAS.500.5106G}, we generated the light curves using differential photometry for all sources in the CCD frames. First, we selected a sample of stars ($\sim$10 stars, same for each night) of similar brightness as the BDs as a reference sample of non-variables, based on the visual inspection of the light curves and their RMS (Root Mean Square of the time-series magnitudes). By averaging these non-variable light curves, we generated an averaged light curve for each night so that individual variations present in each light-curves of the reference stars (if any) get averaged out. Then, subtracting these average magnitudes from our target source magnitudes  (source light curve - average light curve) generates the differential magnitudes. This minimizes the background variations dependent on air mass and instrumental parameters unique to each frame. This subtracted light curve only retains the intrinsic variation of the targeted sources. Thus, the differential photometric technique provides an effective way to detect and classify the variation intrinsic to those variable sources \citep{2018MNRAS.476.2813D, 2019MNRAS.487.1765D, 2021MNRAS.500.5106G}. We have shown the light curves of these BDs from different observing nights in Fig. \ref{lc_plot} which gave a periodic signal in the periodogram analysis \ref{periodogram}. The other non-variable light curves are shown in appendix \ref{non_var_LC} (Fig. \ref{non_var_LC_fig}). The magnitude errors shown in these light curves are calculated by propagating the errors from the instrumental magnitudes' errors estimated in the {\sc phot} task of {\sc IRAF} software\footnote{\url{https://iraf.net/irafhelp.php?val=daophot.phot\&help=Help+Page}}. 
Variability on individual days in the BDs is not significant when compared to the references in the differential photometric data as seen from Fig. \ref{lc_plot}. So if we have any variability that would be of low amplitude. The references show that the RMS of the light curves of BD and references' is similar. However, we proceeded with the periodicity analysis because these BD sources are known variables, to explore any present short-scale dynamics. We investigated individual differential light curves for the presence of any periodic signal. Also, the individual light curves for each source were added together to create a single long-baseline light curve and analysed for any longer periodicities. 

\subsection{Periodogram Analysis: Monte Carlo Simulation}
\label{periodogram}
\begin{figure*}
  \centering
\hspace*{-1.5cm}
\includegraphics[width=1.\linewidth]{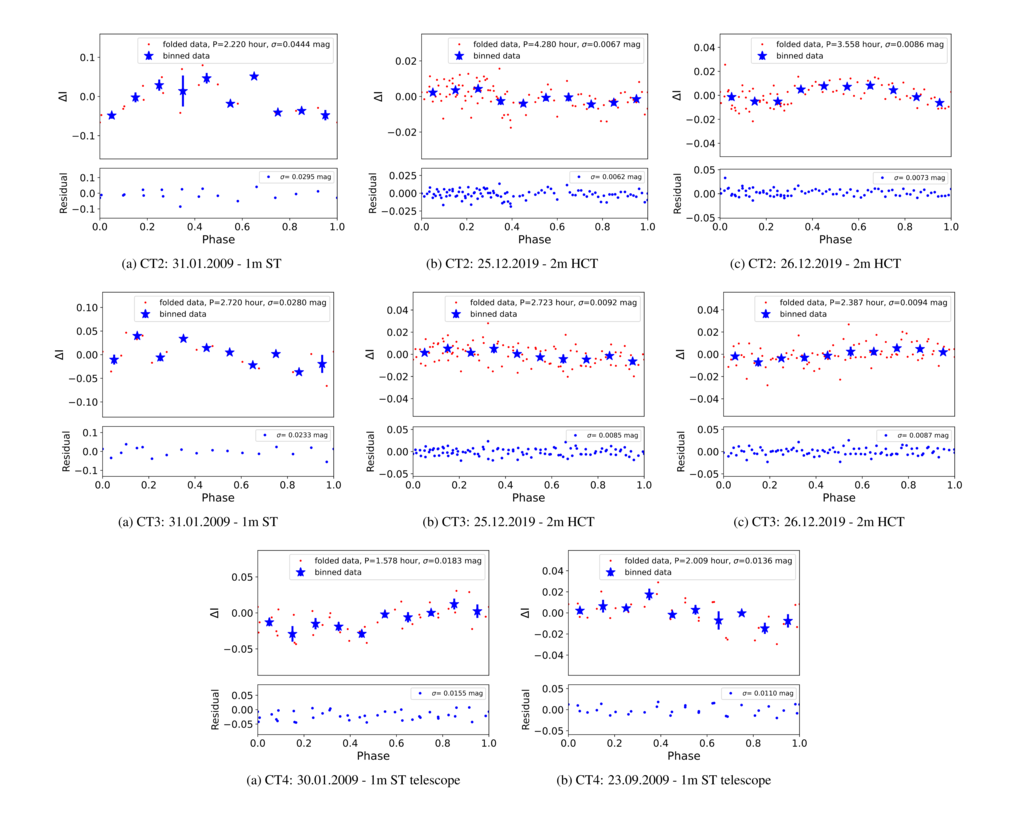}
  \caption{Phase light-curve of sources are shown. Residuals are plotted in the boxes below. The blue stars represent 10-point binned data binned in phase. The source name, dates, and telescopes used to obtain the data are mentioned underneath each panel.}
\label{phase123}
\end{figure*}

For each night, periodograms of observed light curves were computed using NASA Exoplanet Archive Periodogram  Service\footnote{\url{https://exoplanetarchive.ipac.caltech.edu/cgi-bin/Pgram/nph-pgram}} as-well-as using {\it 'astropy.timeseries.LombScargle`} package. It is an algorithm commonly used to find a periodic signal in unevenly spaced data. While computing the LS periodogram, the period values that were the same or near the values of the observing run are discarded as false positives. We divided the total phase (i.e., 0 to 1) into ten bins and averaged the $\Delta I$ in the bin. Phase curves are then formed from the best periods using the most significant peak in the periodogram. We have shown the phase curves of the objects in Fig. \ref{phase123}. The blue stars in these figures are binned in phase data for better visualization and smoothening of the magnitude. The periods were estimated on various individual nights as well as combining the data. The periods and their error in the individual night, RMS of the light curves are mentioned in Table \ref{results-table}. The peak-to-peak amplitudes of light curves, equal to the $2\sqrt{2}\times$RMS, are added in Table \ref{results-table}. The last column contains the amplitude of the sinusoidal fitting using the period from the Monte Carlo simulation. All three BDs showed multiple periodicities on different nights. We found that CT2 has periods ranging from $\sim$2.22 to 4.28 h on individual nights, as shown in Table \ref{results-table}. The CT3 showed periods varying from $\sim$2.39 h to 2.72 hours (Table \ref{results-table}). For CT4, periods were found from 1-m ST data, ranging from $\sim$1.5 to 2.0 hours (Table \ref{results-table}) in two nights. The periodic signals are all found on individual night's time-series data. Combining all observing nights' data for each source, we could not find any new or similar periods. It is to be noted that on consecutive nights' data from 2-m HCT (25th and 26th December 2019), the periods are changing for CT2 and CT3 on a day scale time.

\begin{figure*}
  \centering
\includegraphics[width=0.49\linewidth]{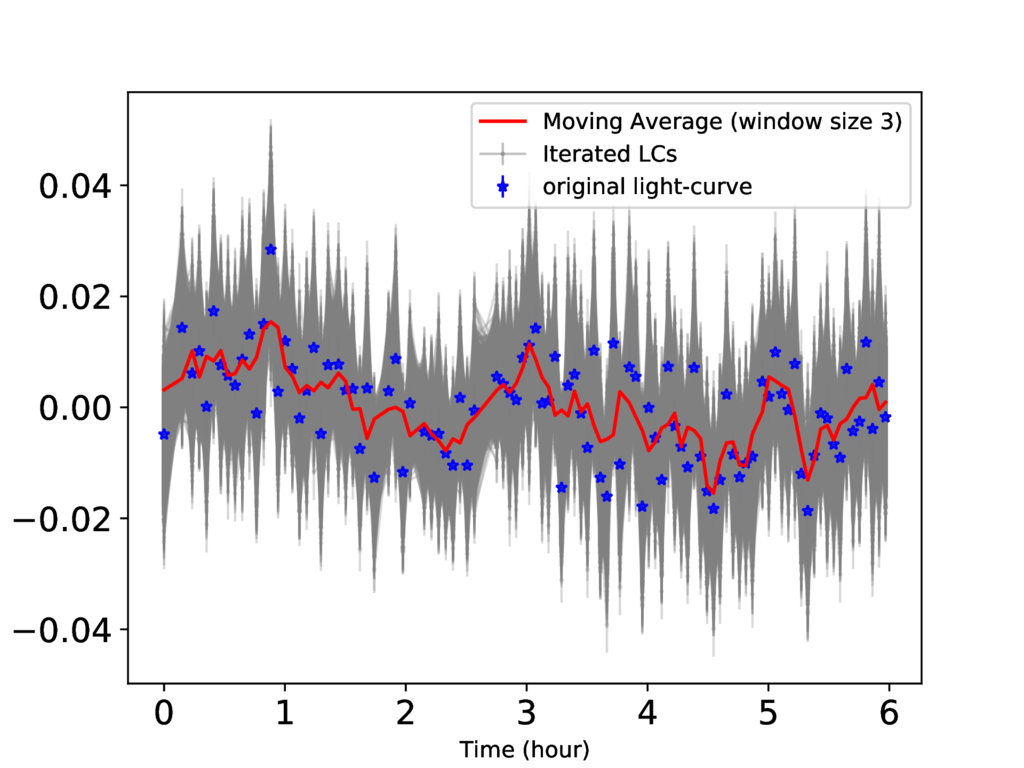}
\includegraphics[width=0.49\linewidth]{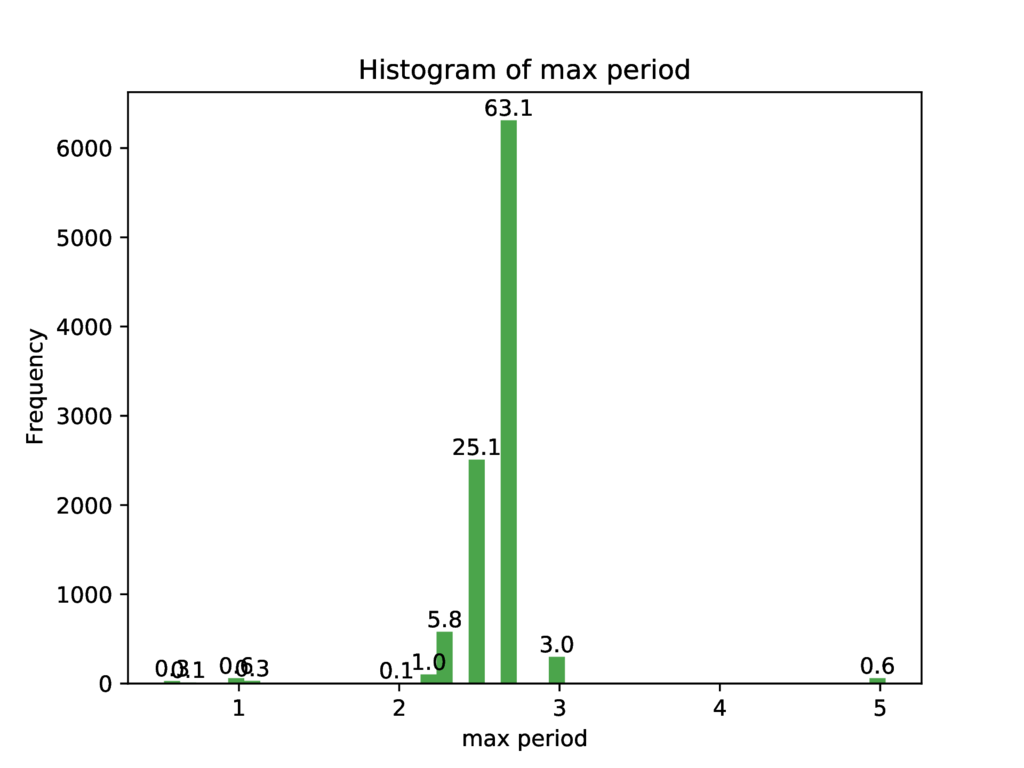}
\includegraphics[width=0.49\linewidth]{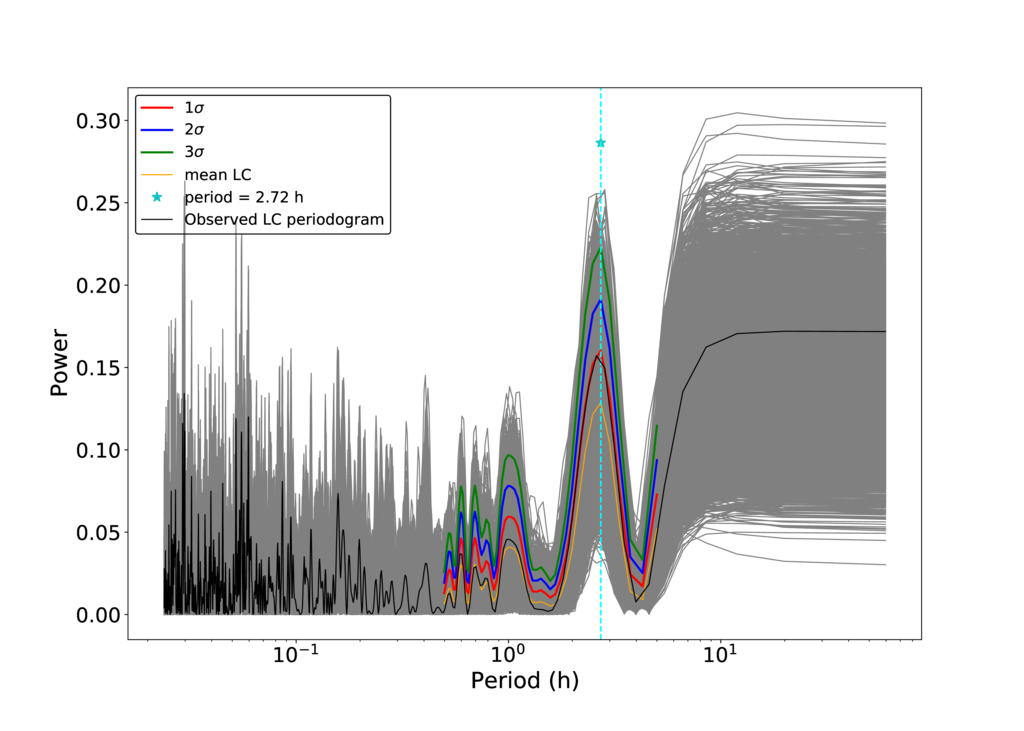}
\includegraphics[width=0.49\linewidth]{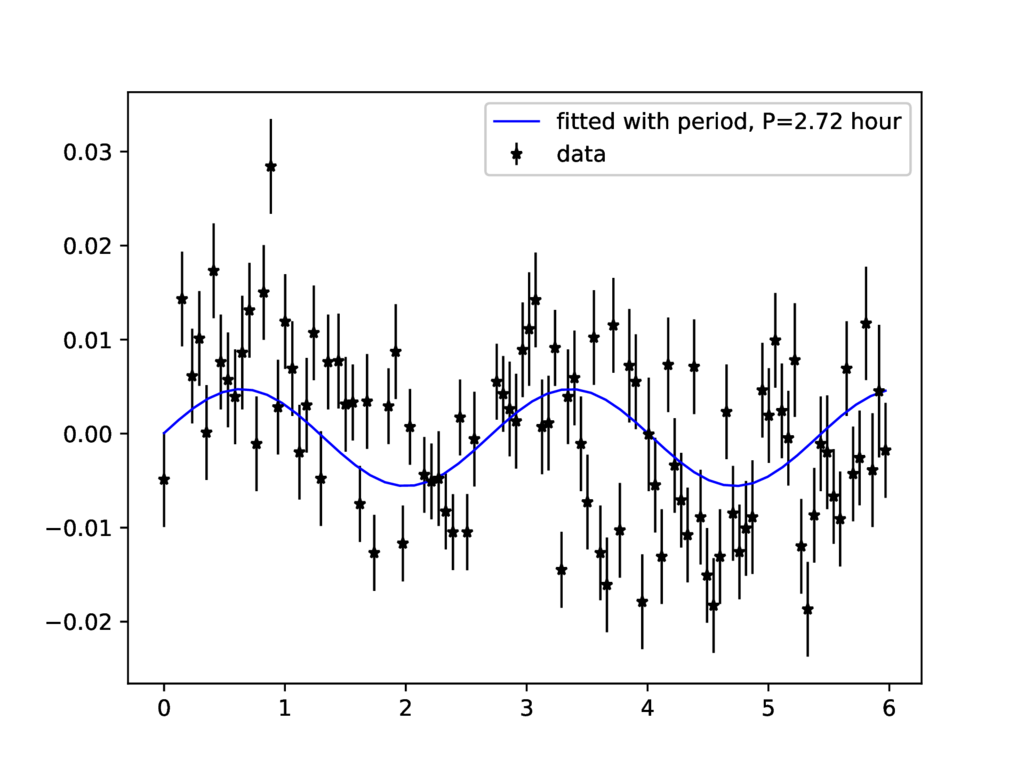}

\caption{Top Left: The 10000 iterated LCs are shown in grey with blue stars being the original LCs. The red line represents a moving average with a window size of 3. Top Right: This gives a visualisation of the period distribution. The highest peak represents the highest number of periods occurring in the simulations. The x-axis is divided in a 0.1 h bin. Bottom Left: The periodogram analysis results of the Monte Carlo Simulated light curves are presented here for CT3 from the December 25, 2019 data. Periodograms of the 10000 synthetic light curves are shown in grey, originating in the Monte Carlo simulation of the original light curve. The black line represents the periodogram of the observed light curve. Red, blue, and green lines represent the one, two, and three $\sigma$ limits of the mean light curve in yellow. The cyan-blue vertical line corresponds to the period found in the LS periodogram of the observed light curve. Bottom Right: The sinusoidal fit to the original LC with the period determined by the MC iteration.}
\label{periodogram1}
\end{figure*}

To comprehensively understand this unstable period i.e., unstable light-curve morphology, we followed the Monte Carlo simulation method used in \cite{2018AJ....155...11M}. For each light curve, we generated 10000 synthetic light curves, slightly different in magnitude values from the original light curve. The original values were redefined using a Gaussian random number generator. We constrain the distribution so that the mean of the 10000 synthetic light curves is the observed light curve and the standard deviation is the photometric uncertainty associated with each point. Lomb-Scargle periodogram (LS periodogram: \citealt{1976Ap...SS..39..447L}, \citealt{1982ApJ...263..835S}, \citealt{2018ApJS..236...16V}) was computed from each source's 10000 synthetic light curve data to find any significant signal present in the light curves using {\it `astropy.timeseries.LombScargle'} package \citep{astropy:2022}. We plotted the synthetic periodograms and calculated the 1$\sigma$, 2$\sigma$, and 3$\sigma$ values (Fig. \ref{periodogram1}). For example, the synthetic periodogram of the Monte Carlo Simulated light curves along with the observed periodogram are presented in Fig. \ref{periodogram1} for CT3 from the data on December 25, 2019. We plotted the mean and 1$\sigma$, 2$\sigma$, and 3$\sigma$ values in yellow, red, blue, and green, respectively. We took the mode of the 10000 values of the period as the periodicity of the data on individual nights, which is closer to the original periods, detected in the observed data. The histogram in the top right panel of Fig. \ref{periodogram1} represents the period distribution. The highest peak represents the periods occurring highest times in the simulations. The x-axis is divided in a 0.1 h bin. This distribution would converge to a Gaussian distribution when n is very large and binning is shorter but that is computationally expensive. 

We plotted the observed light curve in a solid black line, the mean light curve in yellow, and a vertical blue line marking the mean period of the highest peaks. False alarm probability (FAP) measures the probability that a signal with no periodic signal would lead to the detected peak in the LS periodogram. However, red noise complicates both the FAP and detection of peaks \citep{2018ApJS..236...16V}. FAP can't be taken as the authority in the significance of the periodicities in this case. Spread in the periods represents the standard deviation in the periods detected from the synthetic light curves. This quantity would be small for a significant periodic signal. If the uncertainties in the magnitude are high enough, then the synthetic light curves will change significantly and the spread will be higher. We reported the most probable period and spread in the detected period of synthetic light curves (last column of Table \ref{results-table}). The results of the same analysis on other dates are provided as supplementary materials (see appendix section \ref{supl1}, Fig \ref{supCT2}, \ref{supCT3} and \ref{supCT4}).
As given in Table \ref{results-table}, the detected periods have large spreads and taking the FAP into account, we can infer that these short periods are not statistically significant to be considered as true positive signals. This large spread means that the light curve changes its structure for a small change/uncertainty in the data resulting in different periods, which is probably happening here as we got from various date's data. This is also reflected in the fact that the period distribution histogram contains other period peaks though their occurrence is lower. These peaks correspond to signals due to changes in light-curve morphology occurring in the MC iteration.

We have estimated the systemic error, $\delta f = 3 \sigma_N / 2TA \sqrt{N_0}$ ~\citep{1986ApJ...302..757H}, where $\sigma_N^2$ is the variance of the noise in the periodic signal subtracted data, T is the span of data, A is the amplitude of the signal and $N_0$ is the number of independent data points. Using this formula, we estimated the error in the period of the order of $\sim$ 0.01 - 0.1 h for individual nights.

\begin{table*}
\fontsize{7}{7}\selectfont
\caption{Results: Periods and RMS of the variable brown dwarfs from this study}
\label{results-table}
\begin{tabular}{p{1.5cm}p{2.5cm}p{2.cm}p{1cm}p{2.0cm}p{1.0cm}p{2.5cm}p{1.5cm}}
\hline\\
Date  & Target & Period $\pm$ Error$^a$ ~~~~~(h) & RMS (mag) &  Peak-to-peak amplitude ($2\sqrt2~RMS$) & FAP & Period $\pm$ Spread from MC simulation & Amplitude from sine fit (mag) \\ \hline
31.01.2009 	&	CFHT-BD-Tau 2 & 2.22 $\pm$ 0.51 & 0.041 & 0.116 & 0.68  & 2.22 $\pm$ 1.16 & 0.0419\\
08.12.2018 	&	CFHT-BD-Tau 2 & Stochastic & 0.016 & 0.045 & - & - & -\\
01.01.2019  &	CFHT-BD-Tau 2 & Stochastic & 0.056 & 0.158  & -  & - &-	\\
02.01.2019  &	CFHT-BD-Tau 2	& Stochastic & 0.051 & 0.144  & - & - &-\\
25.12.2019  &   CFHT-BD-Tau 2 &  4.28 $\pm$ 0.07 & 0.007 & 0.020 & 0.72  & 4.28 $\pm$ 1.66 & 0.0036\\
26.12.2019  &   CFHT-BD-Tau 2 & 3.56 $\pm$ 0.30 & 0.009 & 0.025 & 0.00005  & 3.48 $\pm$ 1.23 & 0.0067 \\\hline
31.01.2009 	&	CFHT-BD-Tau 3 & 2.72 $\pm$0.08 & 0.027 & 0.076 & 0.60 & 3.07 $\pm$  1.22 & 0.0216\\
08.12.2018 	&	CFHT-BD-Tau 3 &Stochastic & 0.017 & 0.048  & - & -&- \\
01.01.2019  &	CFHT-BD-Tau 3	&Stochastic& 0.044 & 0.124 & - & - &- \\
02.01.2019  &	CFHT-BD-Tau 3	& Stochastic & 0.044 & 0.124 & - & - &- \\
25.12.2019  &   CFHT-BD-Tau 3 &2.72 $\pm$ 0.31& 0.009 & 0.025 & 0.04 & 2.72 $\pm$ 1.32 & 0.0052 \\ 
26.12.2019  &   CFHT-BD-Tau 3 & 2.39 $\pm$ 0.99 & 0.009 & 0.025 & 0.05 & 2.17 $\pm$ 1.59 & 0.0048 \\\hline
30.01.2009  &	CFHT-BD-Tau 4	& 1.58 $\pm$ 0.92 & 0.031 & 0.088 & 0.003  & 1.63 $\pm$ 1.40 & 0.0137	\\
23.10.2009	&	CFHT-BD-Tau 4 & 2.03 $\pm$ 0.86 & 0.020 & 0.057 & 0.36  & 2.03 $\pm$ 1.38 & 0.0177\\
09.12.2018  &	CFHT-BD-Tau 4 & Stochastic & 0.080 & 0.226  & -  & - & - \\\hline

\end{tabular}

$^a$The estimated periods have systematic errors, which is discussed in section \ref{periodogram}.

\end{table*}

\subsection{Spectral Energy Distribution of the three BDs}
\label{flare_section}
\begin{figure}[htbp]
    \centering
    \begin{subfigure}[b]{0.32\linewidth}
        \includegraphics[width=\linewidth]{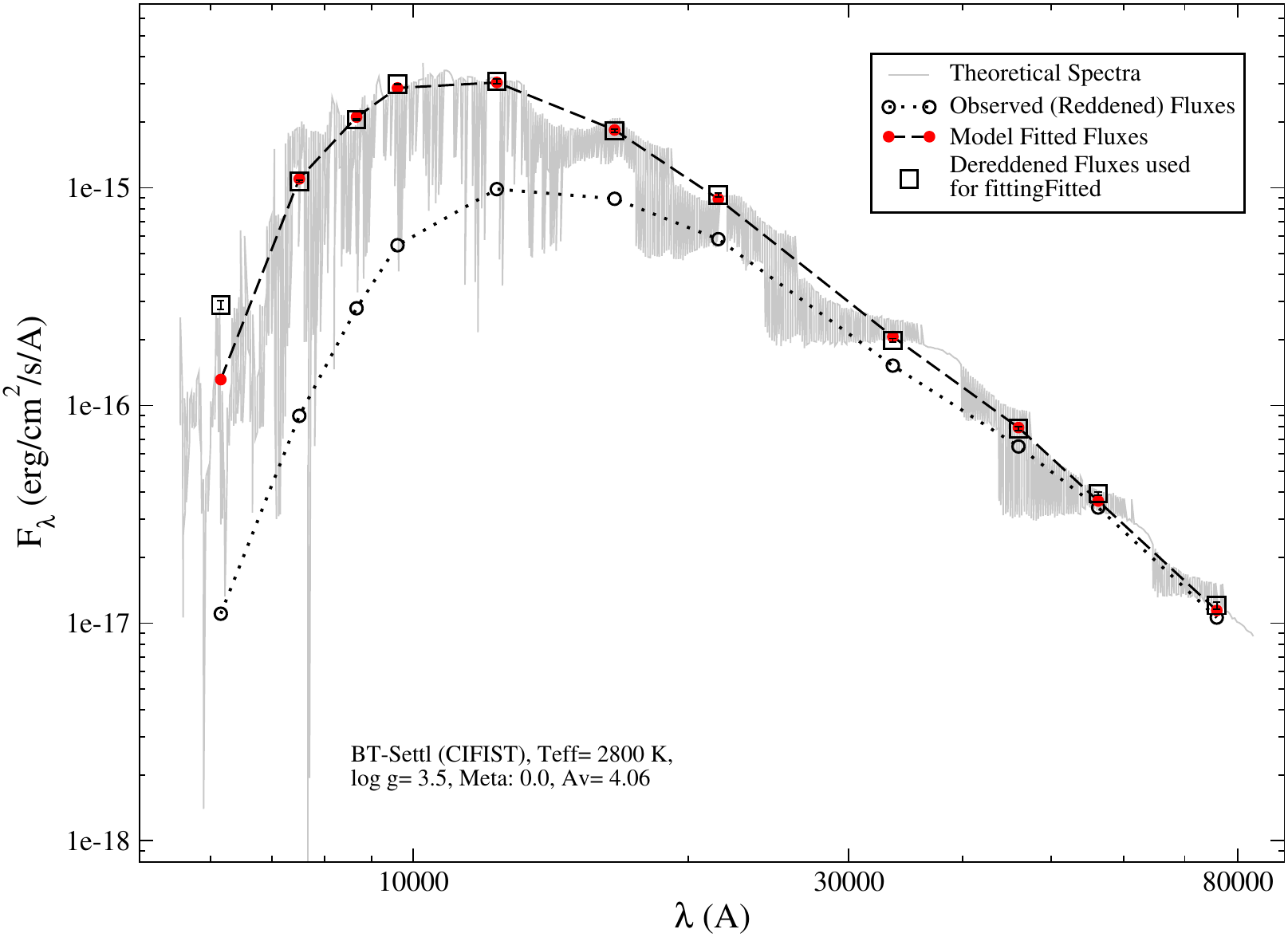}
        \caption{CT2}
        \label{fig:sub1}
    \end{subfigure}
    \begin{subfigure}[b]{0.32\linewidth}
        \includegraphics[width=\linewidth]{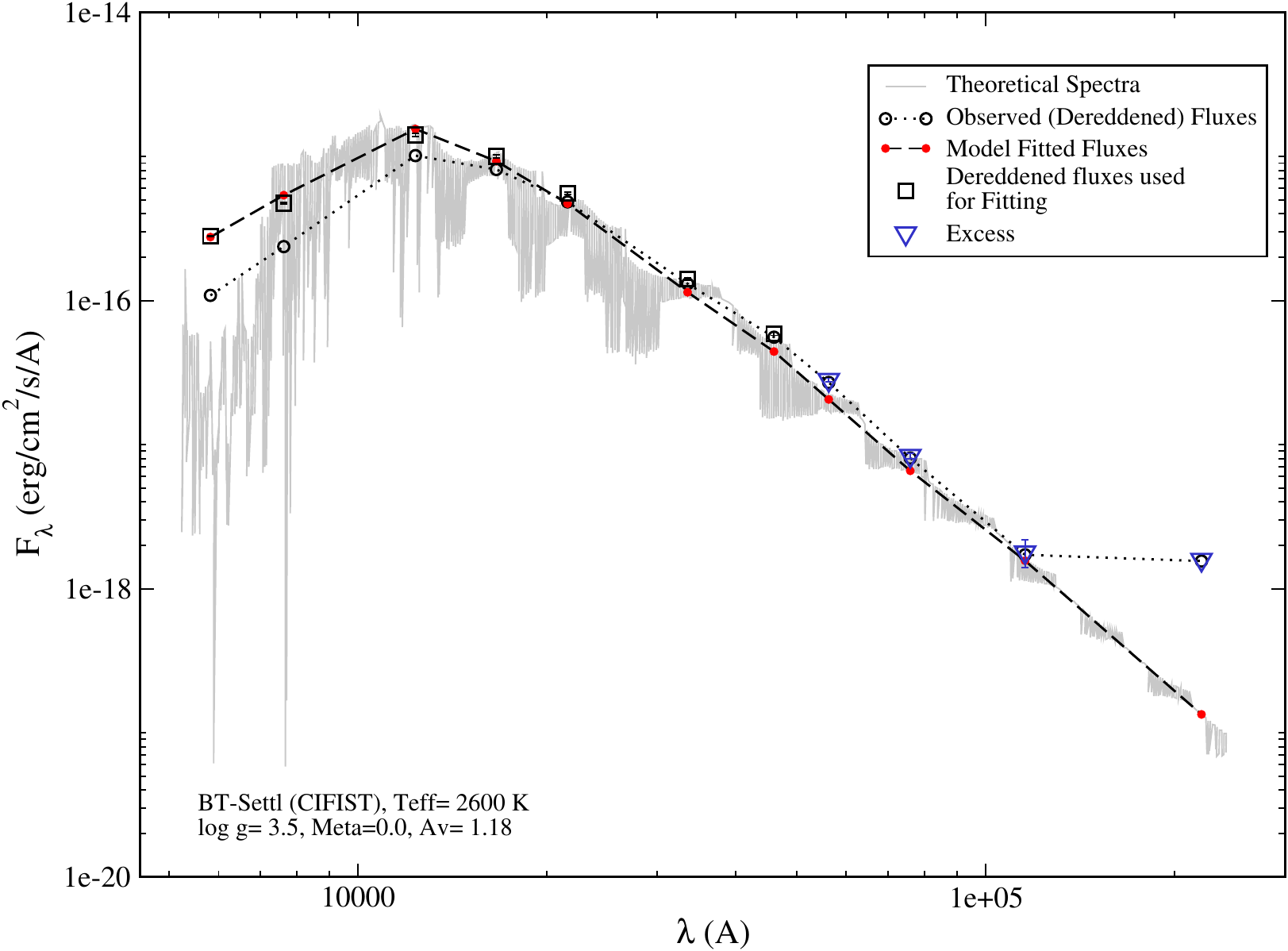}
        \caption{CT3}
        \label{fig:sub2}
    \end{subfigure}
    \begin{subfigure}[b]{0.32\linewidth}
        \includegraphics[width=\linewidth]{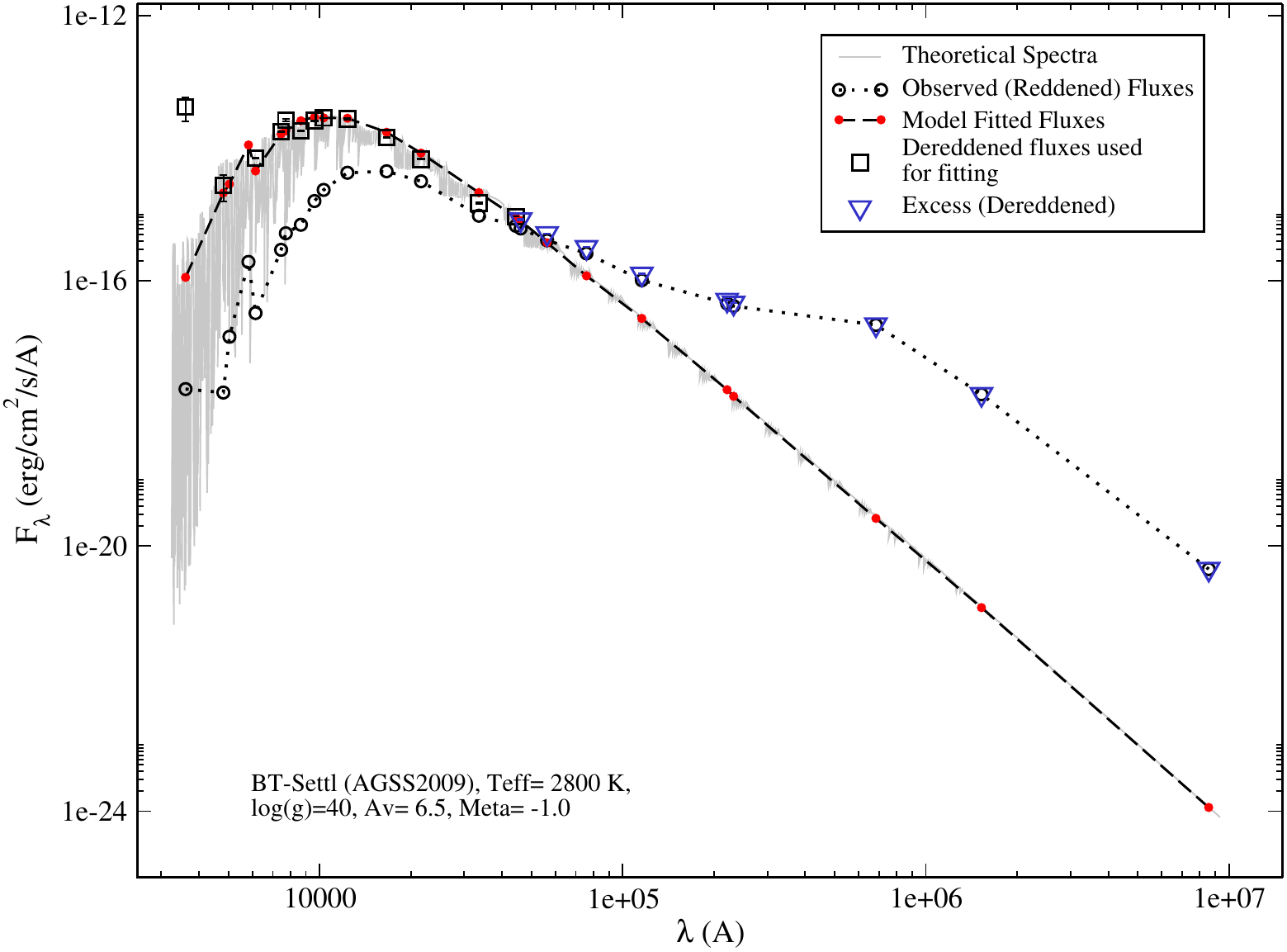}
        \caption{CT4}
        \label{fig:sub3}
    \end{subfigure}
    \caption{Spectral Energy Distributions (SED) of the three BDs: CT2, CT3 and CT4 (from left to right) are constructed using the models from \citep{2009ARA&A..47..481A, 2011SoPh..268..255C, 2013MSAIS..24..128A}. The black dotted line with empty black circles represents the input/observed fluxes, empty black boxes represent the dereddened input data used for model fitting, and the red solid circles with dashed black lines represent fitted output fluxes. The grey line indicates the BT-Settl model spectra with T$_{eff}$, $log(g)$ and other parameters taken from SED fitting respectively. The inverted blue triangles represent the infrared excess corresponding to each observed flux which is not used in the fit.}
    \label{flare_SED}
\end{figure}

We constructed the Spectral Energy Distribution (SED: Fig. \ref{flare_SED}) for BDs using the interactive service ``Virtual Observatory SED Analyzer" (VOSA\footnote{\url{http://svo2.cab.inta-csic.es/svo/theory/vosa/index.php}}). VOSA tool provides physical properties like luminosity, temperature, and radius of any source by fitting model synthetic spectra or photometric fluxes to multi-wavelength observed data provided by the user or gathered automatically from available catalogues using the ``VO photometry" tools \citep{2008A&A...492..277B, refId1}.\\

In constructing the Spectral Energy Distribution (SED) of CT2, we incorporated data from the PAN-STARRS r, i, z, y band \citep{chambers2019panstarrs1surveys}; near-IR ($JHK$) data from 2MASS and WISE W1 and W2 band data; and IRAC 5.8 $\mu m$ and 8.0 $\mu m$ data (Infrared Array Camera; \citealp{2004ApJS..154...10F}) using the VO photometry tool.\\ 

For the SED construction of CT3, we utilized data from GAIA G, Grp band \citep{2016A&A...595A...1G,2023A&A...674A...1G}; near-IR ($JHK$) data from 2MASS point source catalogue \citep{2003tmc..book.....C} and W1, W2, W2 and W3 data from Wide-field Infrared Survey Explorer (WISE) \citep{Wright_2010, 2012wise.rept....1C}.\\

In constructing the SED of CT4, we used near-IR ($JHK$) data from 2MASS point source catalogue, $u$, $r$, $i$ band data from SDSS DR12 (Sloan Digital Sky Survey-  \citealp{2015ApJS..219...12A}), g, z, y band data from PAN-STARRS, GAIA Grp band data, W1, W2, W2 and W3 data from WISE, IRAC 3.6 $\mu m$, 4.5 $\mu m$, 5.8 $\mu m$ and 8.0 $\mu m$ data, Y band data from UKIDSS, MIPS 24 $\mu m$, 70 $\mu m$ (Mid-Infrared Photometer for Spitzer; \citealt{2014ApJ...784..126E}) from Spitzer survey of young stellar clusters \citep{2006ApJ...647.1180L, 2010ApJS..186..259R} and sub-mm 850 $\mu m$ data \citep{2003ApJ...593L..57K} from JCMT's (James Clerk Maxwell Telescope) sub-millimeter common-User bolometer array (SCUBA; \citealp{1999MNRAS.303..659H}). Applying BT-Settl model \citep{2011ASPC..448...91A, 2012RSPTA.370.2765A} with \cite{2009ARA&A..47..481A} abundances (hereafter, BT-settl (AGSS2009)) and with \citep{2011SoPh..268..255C} abundances (hereafter, BT-settl (AGSS2009)) we find the effective temperature luminosity, mass and radius of these three BDs (see Table. \ref{SED_table}). This luminosity and temperature from the SED are in good agreement as previously reported for these young BDs \citep{Rebull_2010, 2018ApJ...858...41Z}.

A radius of 0.65$R_{\odot}$ and luminosity of $\sim$0.01 - 0.027$L_{\odot}$ are reported in the literature \citep{2018ApJ...861...76P, 2019ApJ...872..158A} for CT4. The present estimation of radius and luminosity could be more robust as we used a wide range of fluxes, from optical (SDSS) to sub-mm (850 microns: JCMT) to fit the SED. 

\begin{table*}
    \centering
    \caption{Results from SED fitting}
    \label{SED_table}
    \begin{tabular}{ p{3cm} p{3.0cm} p{3.0cm} p{3.2cm} p{3.0cm}}
    \hline
        Parameter & Value & Value & Value & Reference  \\\hline    
        Object & CFHT-BD-Tau 2 & CFHT-BD-Tau 3 & CFHT-BD-Tau 4 & -\\
        SpT & M7.5 & M7.75 & M7 & M01 \\
        T$_{\rm eff}$ [K]   & 2700$\pm$50 & 2600$\pm$50  & 2800$\pm$50  & This work \\
        $\rm log~g $ [$g/cm^2$]  & 3.5$\pm$0.25 & 3.5$\pm$0.25 &  4.00$\pm$0.25 & '' \\
        Mass [$M_{\odot}$] & 0.047$\pm$0.030 & 0.030$\pm$0.020 & 0.066$\pm$0.040 & '' \\
        $A_v$ & 4.06$\pm$0.13 & 1.18$\pm$0.03 & 6.37 & P18  \\
        R/$R_{\odot}$ & 0.64$\pm$0.03 & 0.52$\pm$0.02 & 0.42$\pm$0.04& This work	\\
        $L/L_{\odot}$ & 0.023$\pm$0.001 & 0.011$\pm$0.001 &0.010$\pm$0.002 & '' \\
        $T_{flare}$ [k] &-&-& 10000 & '' \\
        SED Model & BT-Settl (CIFIST) & BT-Settl (CIFIST)& BT-Settl (AGSS2009) & \\
        \hline\\
\end{tabular}\\
    M01: \cite{2001ApJ...561L.195M};  P18: \citep{2018ApJ...861...76P}. 
\end{table*}

In the following sections, we have individually discussed the results of the three BD sources from our study.

\subsection{CFHT-BD-Tau 2}

For CT2, hour-scale periodicity was detected in 3 out of 6 nights from ground-based data; others were non-periodic or stochastic. We did not find significant periods in the combined time-series data for this source. The period varies from $\sim$2.2 to $\sim$4.3 hours. And from the Monte-Carlo simulation, we get similar periodicity. However, the spread is too high to infer any insights into this object. TESS data were not available for this data till now, but data from Kepler had already confirmed a day scale period ($\sim$ 2.9 d) for CT2, which is rotating faster than the other two BDs. However, the properties like mass, radius, SpT, brightness etc. are similar to the other two BDs, especially the CT3. We constructed the SED of this source (Fig. \ref{flare_SED}) and re-evaluated its mass, radius, temperature etc. (Table. \ref{table:SED}) which matches with previous estimation within error. From the SED, it is inferred that the source has low to very low infrared excess. High-precision data like TESS photometry and other photometric as well as spectroscopic data along with surface modeling could probe deeper into the atmosphere of this relatively faster rotating BD.

\subsection{CFHT-BD-Tau 3}
We normalized the PDCSAP light curve of CT4 after removing the outliers from the light curve. The corrected light curves are shown in Fig. \ref{tess_lc_ct443}. Lomb-Scargle periodogram analysis \citep{1976Ap...SS..39..447L,1982ApJ...263..835S} was performed on the corrected light curves to generate phase curves with the most significant period. We removed the flare events from the light curves before the periodogram analysis. The periodogram gives a sharp peak at 0.97 days in sector 70 and 0.96 days in sector 71 (Fig. \ref{tess_phase_ct2}). This value matches well with the previously reported period for this object (\citealt{Scholz_2018}, \citealt{2020AJ....159..273R}). The median count rate, RMS of the light curve, and results are mentioned in Table \ref{results-table_tess}. We folded the light curve with this period to generate phase curves for CT3 (Fig. \ref{tess_phase_ct2} right panels). We showed the folded data in blue dots and binned the data (in black stars) for better visualization. 

From ground-based data, hour-scale periodicity was detected in 3 out of 6 nights; others were non-periodic or stochastic. We did not find significant periods in the combined time-series data for this source. The period varies from $\sim$2.4 to 2.7 hours. And from the Monte-Carlo simulation, we get similar periodicity. However, the spread is too high to infer any insights into this object. TESS data were not available for this data till now, but data from Kepler had already confirmed a day scale period ($\sim$ 0.96 d) for CT3, which is rotating faster than the other two BDs. However, the properties like mass, radius, SpT, brightness etc. are similar to the other two BDs, especially the CT2. High-precision data like TESS photometry and other photometric as well as spectroscopic data along with surface modeling could probe deeper into the atmosphere of this relatively faster rotating BD.

We constructed the SED of this source (Fig. \ref{flare_SED}) and re-evaluated its mass, radius, temperature etc. (Table. \ref{table:SED}) which also matches with previous estimation within error. For this object too, the SED showed more infrared excess than CT2 and less than CT4 which is also reflected in the $(J - K)$ colours in Fig. \ref{CMD}.

\subsection{CFHT-BD-Tau 4}	
\subsubsection{Photometric variability}
\label{TESS_photometry}

For CT4, periods were found ranging from $\sim$1.5 to 2.0 hours (Table 3) in two nights from ground-based data in 2009. The periodic signals are all found on individual night’s time-series data. Combining all observing nights’ data for each source, we could not find any new or similar periods. From the Monte-Carlo simulation, we get similar periodicity with the spread in period too high to infer any insights.

\begin{figure}
\begin{minipage}[b]{1\linewidth}

  \centering
      \subcaptionbox{}[.49\linewidth][c]{%
    \includegraphics[width=.98\linewidth]{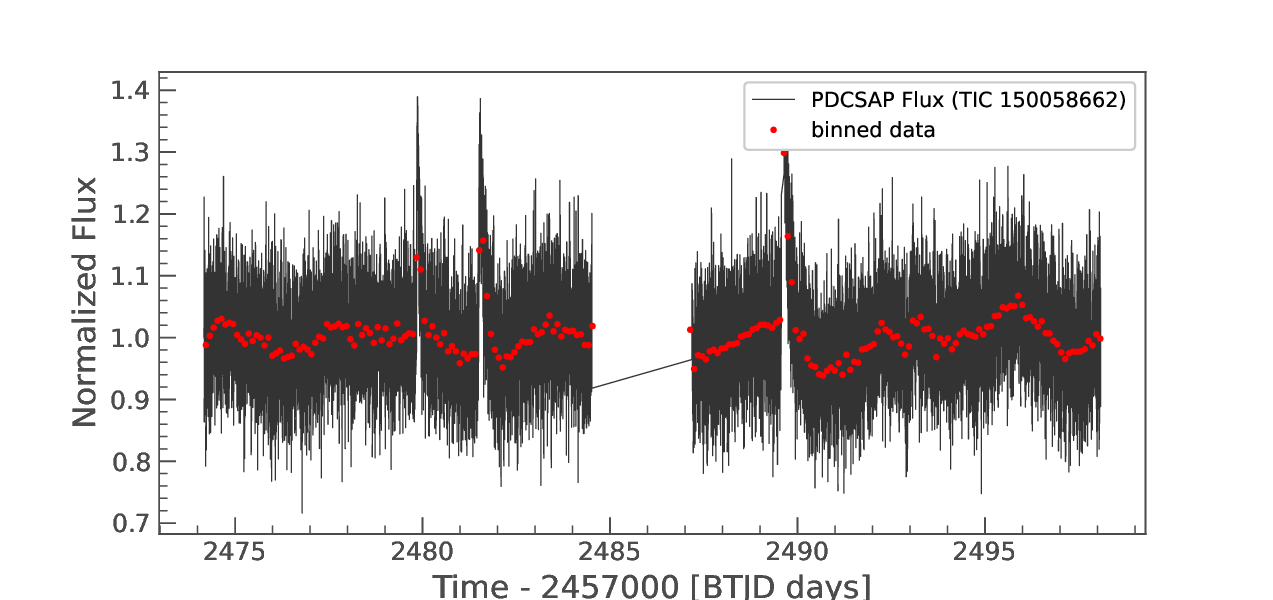}}
      \subcaptionbox{}[.49\linewidth][c]{%
    \includegraphics[width=.98\linewidth]{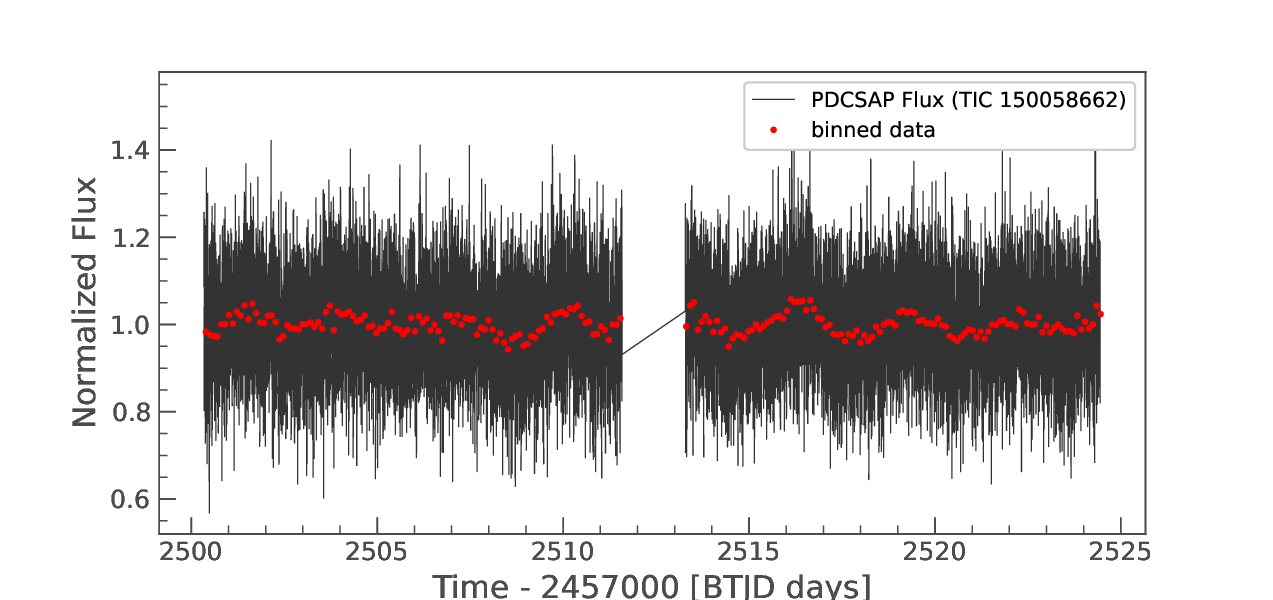}} \quad
          \subcaptionbox{}[.49\linewidth][c]{%
    \includegraphics[width=.98\linewidth]{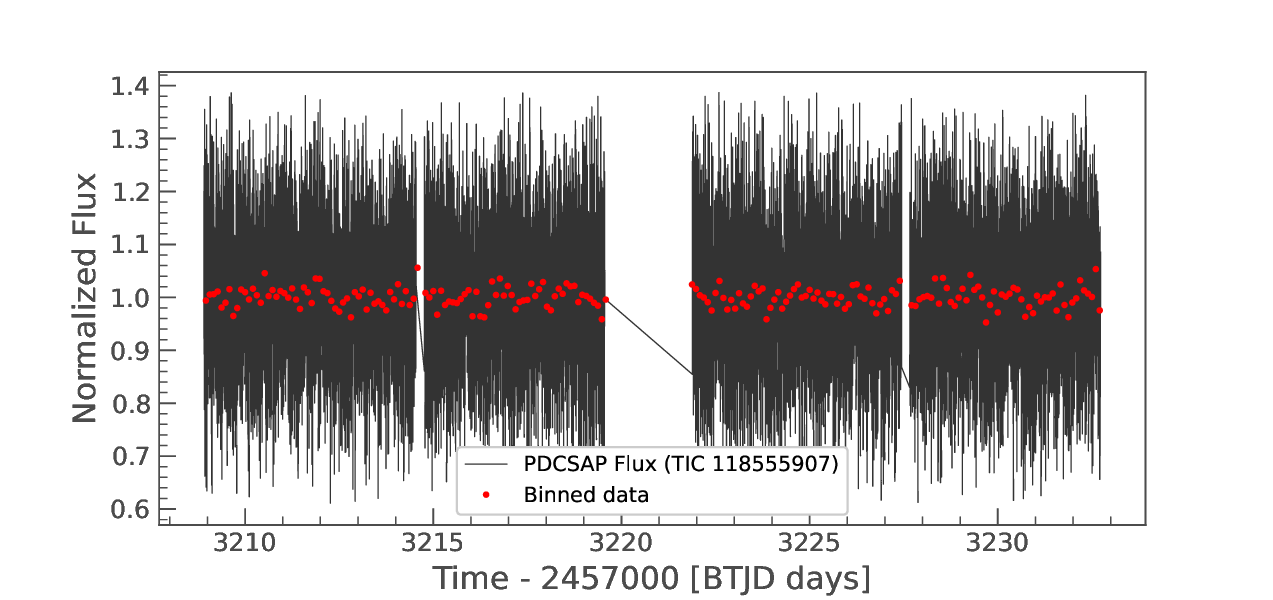}}
      \subcaptionbox{}[.49\linewidth][c]{%
    \includegraphics[width=.98\linewidth]{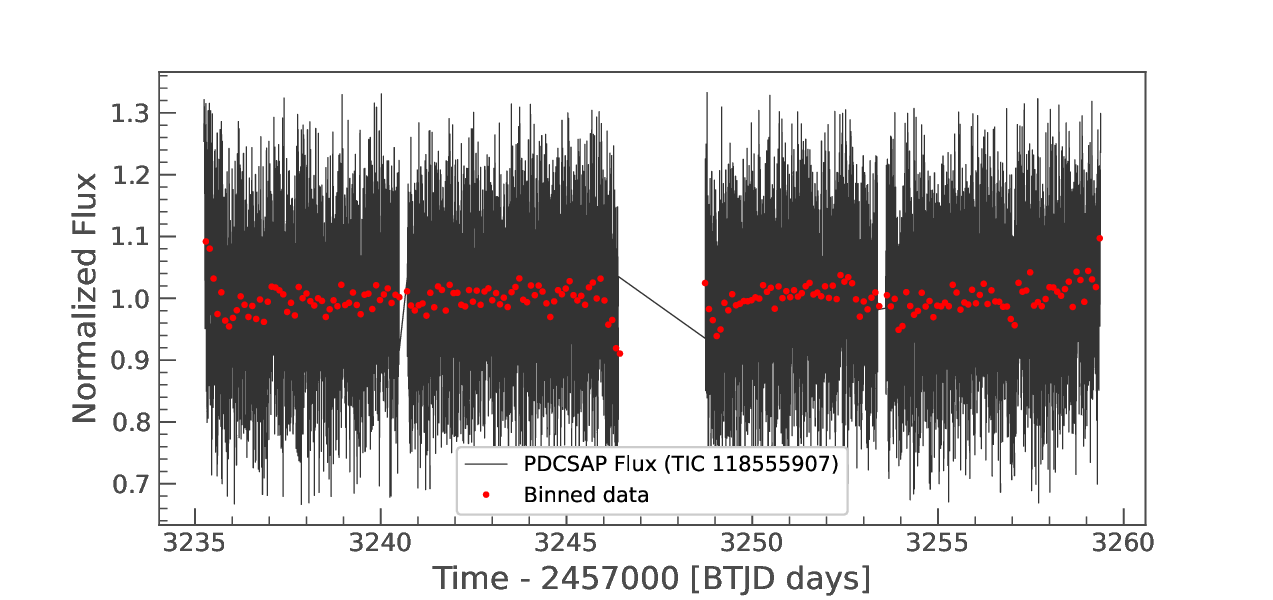}} \quad

\end{minipage}
 \caption{(a,b.) The normalized PDCSAP light curve of CFHT-BD-Tau 4 from 2 min cadence data of sectors 43 and 44 are shown here. (c,d.) The normalized PDCSAP light curves of CFHT-BD-Tau 2 from 2 min cadence data of sectors 70 and 71 are shown here. The x-axis is the time in TESS Barycentric Julian Days (BTJD) and the y-axis is the normalized TESS flux ($e^{-}$/s). The light curve is corrected for outliers and normalized. Red points show 150 min binned data for better visualization.}	
   \label{tess_lc_ct443}
\end{figure}

From TESS data, we normalized the PDCSAP light curve of CT4 after removing the outliers from the light curve similarly. The corrected light curves are shown in Fig. \ref{tess_lc_ct443}. Lomb-Scargle periodogram analysis \citep{1976Ap...SS..39..447L,1982ApJ...263..835S} was performed on the corrected light curves to generate phase curves with the most significant period. We removed the flare events from the light curves before the periodogram analysis.

\begin{figure*}
\begin{minipage}[b]{1\linewidth}

  \centering
    \subcaptionbox{}[.49\linewidth][c]{%
        \includegraphics[width=.98\linewidth]{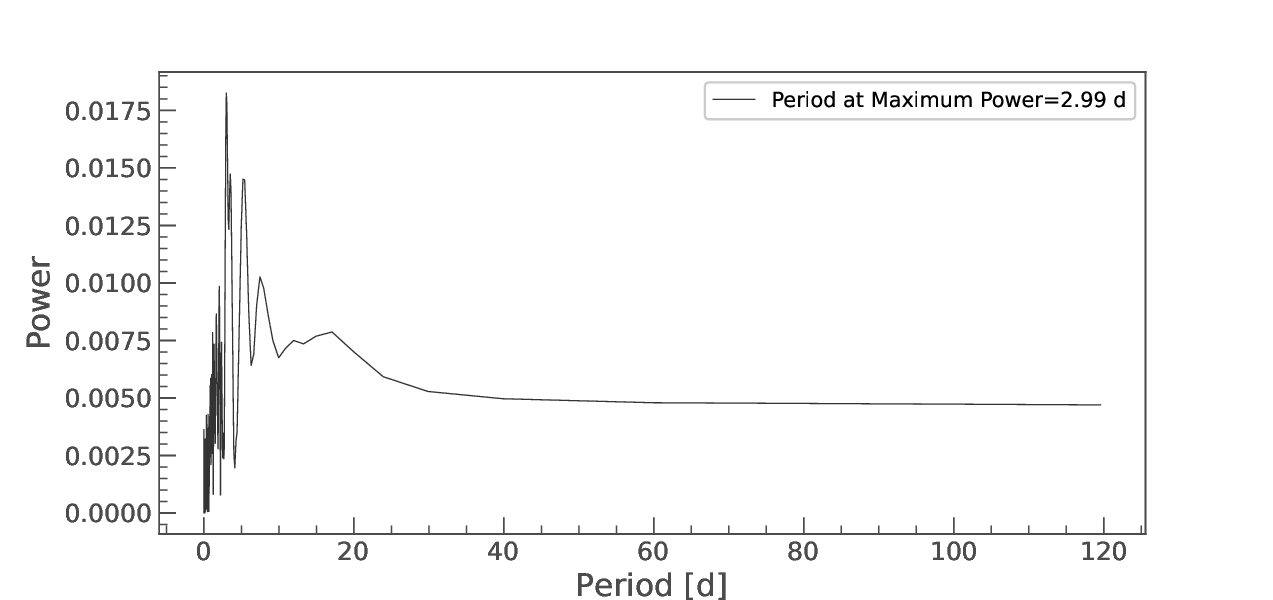}}
    \subcaptionbox{}[.49\linewidth][c]{%
        \includegraphics[width=.98\linewidth]{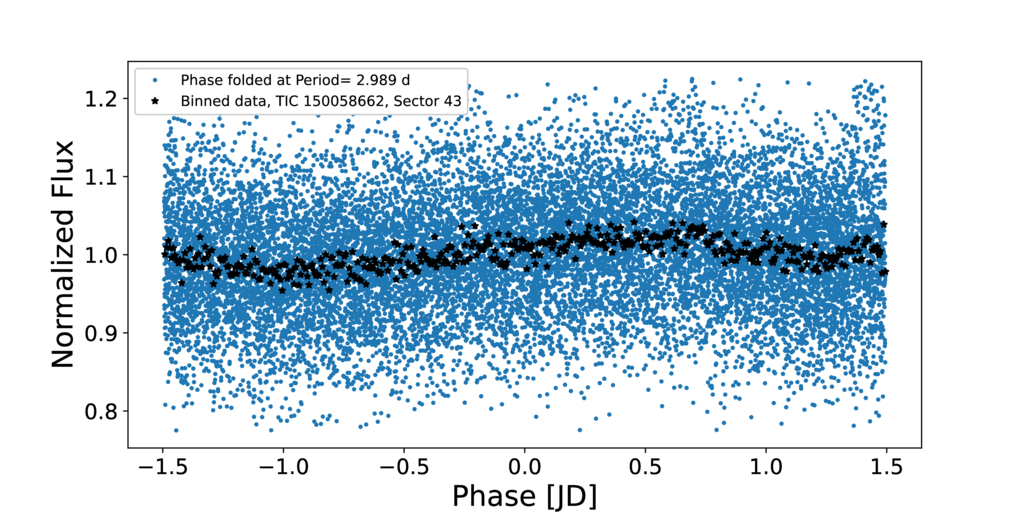}}
    \subcaptionbox{}[.49\linewidth][c]{%
        \includegraphics[width=.98\linewidth]{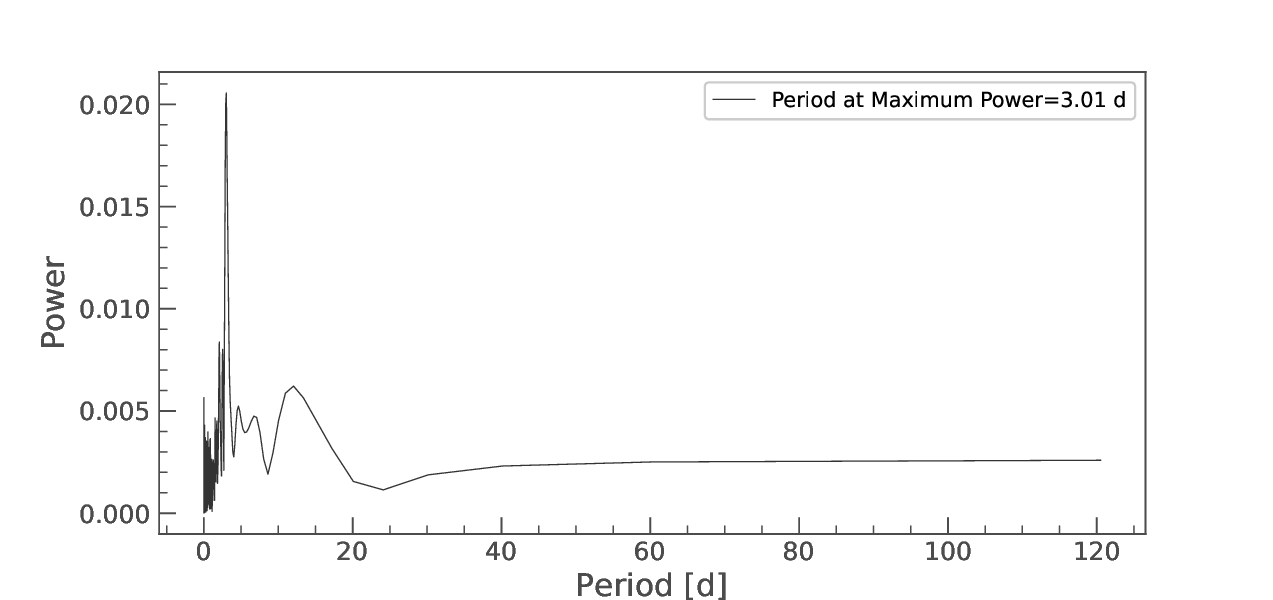}}
    \subcaptionbox{}[.49\linewidth][c]{%
        \includegraphics[width=.98\linewidth]{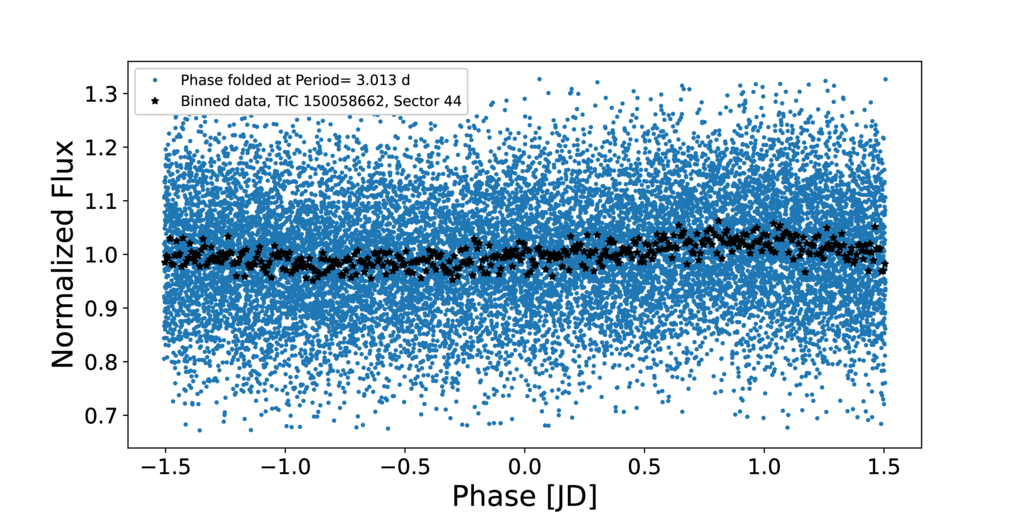}}

\end{minipage}
 \caption{(a,c.) Periodogram for sectors 43 and 44 of PDCSAP light curve with y-axis showing power and x-axis showing period in days. The periodograms give a sharp peak at 2.99 and 3.01 days. (b,d.) Phase curve of CFHT-BD-Tau 4 folded with their respective periods. The blue dots show the phase curve, and the black stars are 150-point binned data for better visualization of the periodic nature. The x-axis shows phase, and the y-axis shows flux in $e^{-}/s$.}	
   \label{tess_phase_ct443}
\end{figure*}

\begin{figure*}
\begin{minipage}[b]{1\linewidth}
  \centering
      \subcaptionbox{}[.49\linewidth][c]{%
    \includegraphics[width=.98\linewidth]{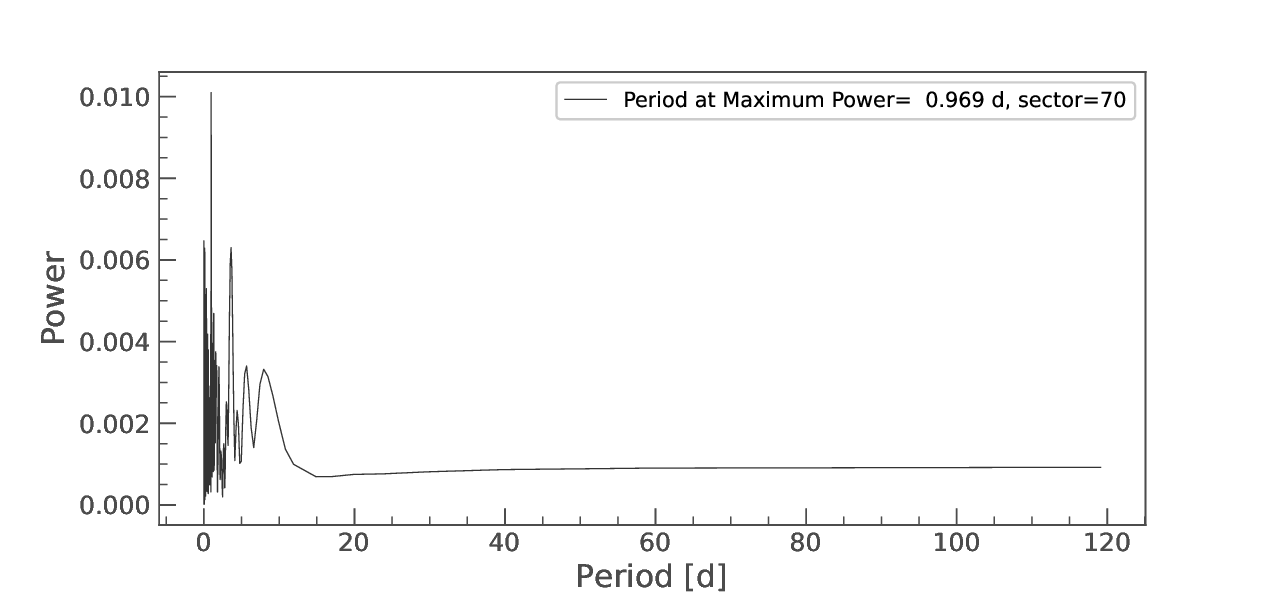}}
      \subcaptionbox{}[.49\linewidth][c]{%
    \includegraphics[width=.98\linewidth]{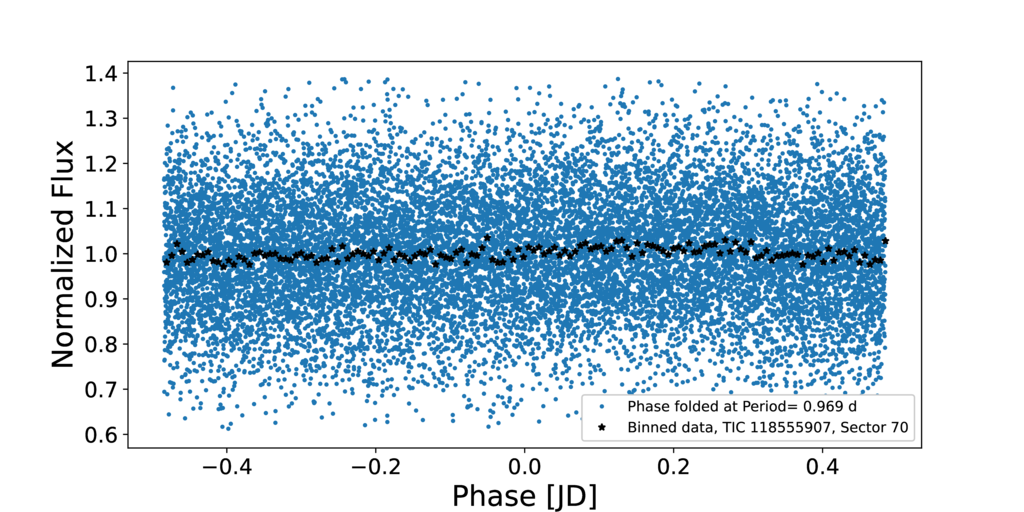}} \quad
      \subcaptionbox{}[.49\linewidth][c]{%
    \includegraphics[width=.98\linewidth]{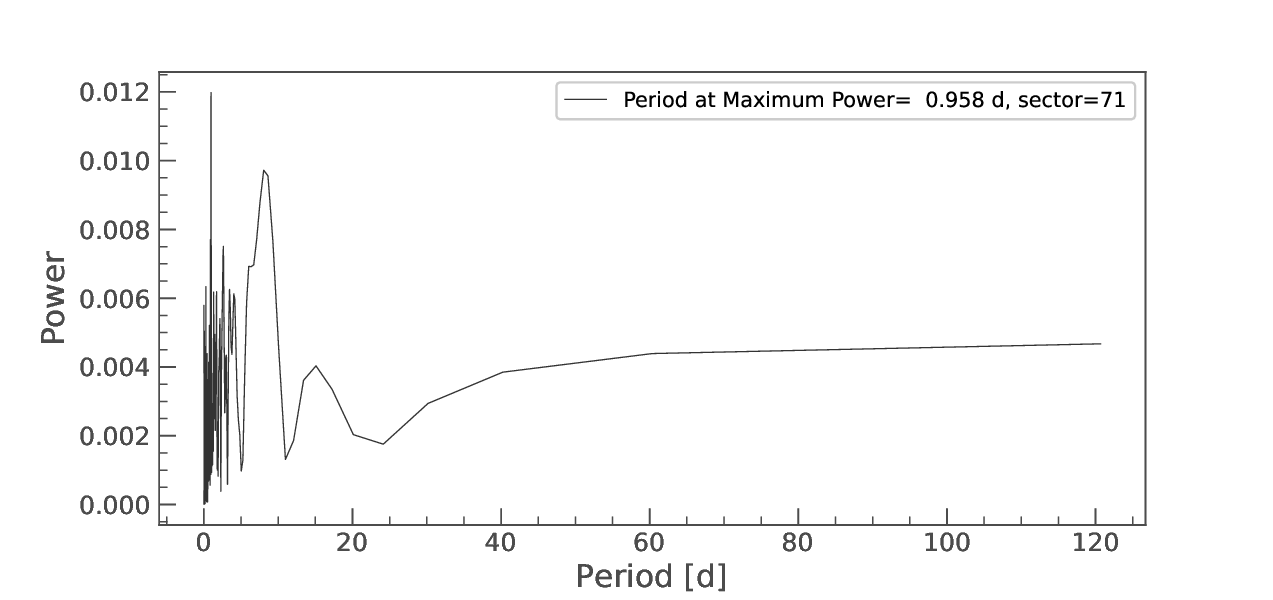}}
      \subcaptionbox{}[.49\linewidth][c]{%
    \includegraphics[width=.98\linewidth]{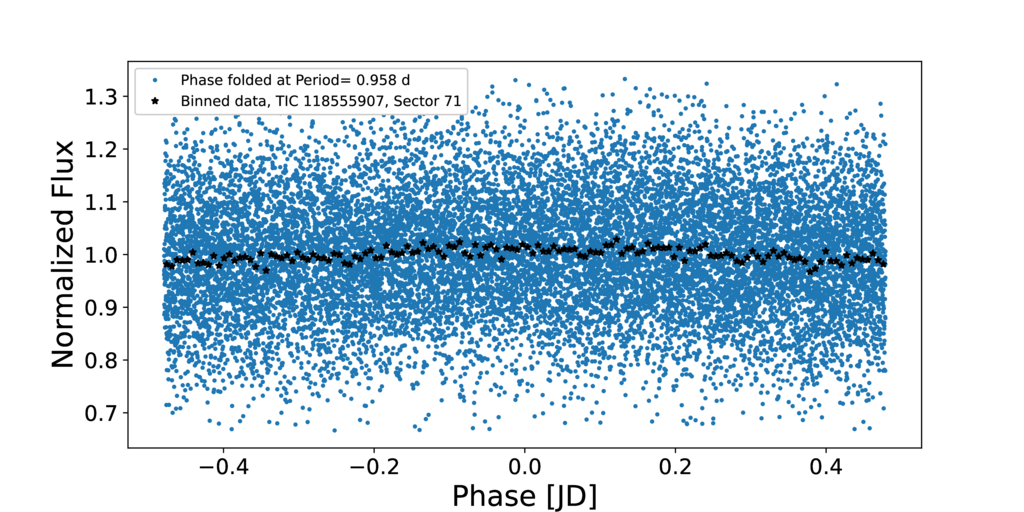}} \quad
\end{minipage}
 \caption{(a,c.) Periodograms for CT3 in the sector 70 and 71 PDCSAP light curve with the y-axis showing power and x-axis showing period in days are shown. The periodograms give a sharp peak at 0.96 and 0.97 days. (b,d.) Phase light curves of CT3 folded with respective periods. The blue dots show the phase curve, and the black stars are binned data for better visualization of the periodic nature. The x-axis shows phase, and the y-axis shows flux in $e^{-}/s$.}	
   \label{tess_phase_ct2}
\end{figure*}

We found a significant periodic variation in the LS periodogram of CT4, which gives a sharp peak at 2.99 days in sector 43 and 3.01 days in sector 44 (left panels; Fig. \ref{tess_phase_ct443}). This value matches well with the previously reported period for this object (\citealt{Scholz_2018}, \citealt{2020AJ....159..273R}). The median count rate, RMS of the light curve, and results are mentioned in Table \ref{results-table_tess}. We folded the light curve with this period to generate phase curves for CT4 (right panels; Fig. \ref{tess_phase_ct443}). We showed the folded data in blue dots and binned the data (in black stars) for better visualization with {\it  PyAstronomy.pyasl.binningx0dt}\footnote{\url{https://pyastronomy.readthedocs.io/en/latest/pyaslDoc/aslDoc/binning.html}} \citep{pya} reducing the original points by 500. These are shown in black stars.

\begin{table*}
\caption{Period of CT2 and CT4 from TESS data}
\label{results-table_tess}
\hspace{1.0cm}
\begin{tabular}{p{2.cm}p{3.7cm}p{2.0cm}p{2.2cm}p{1.7cm}}
\hline
TESS sector & Name / Target	 & Period (day) & Median count of LC [$e^{-}/s$]& RMS  [$e^{-}/s$]\\ \hline
43 	&    CT4 / TIC 150058662 & 2.99 & 6.12 & 0.21 \\
44 	&	 CT4 / TIC 150058662 & 3.01 & 5.91 & 0.24 \\
70  &    CT2 / TIC 118555907 & 0.97 & 6.39 & 0.24 \\
71  &    CT2 / TIC 118555907 & 0.96 & 6.08 & 0.25 \\\hline
\end{tabular}
\end{table*}

\begin{figure*}
\begin{minipage}[b]{0.9\linewidth}

  \centering
    \subcaptionbox{}[.49\linewidth][c]{%
    \includegraphics[width=.98\linewidth]{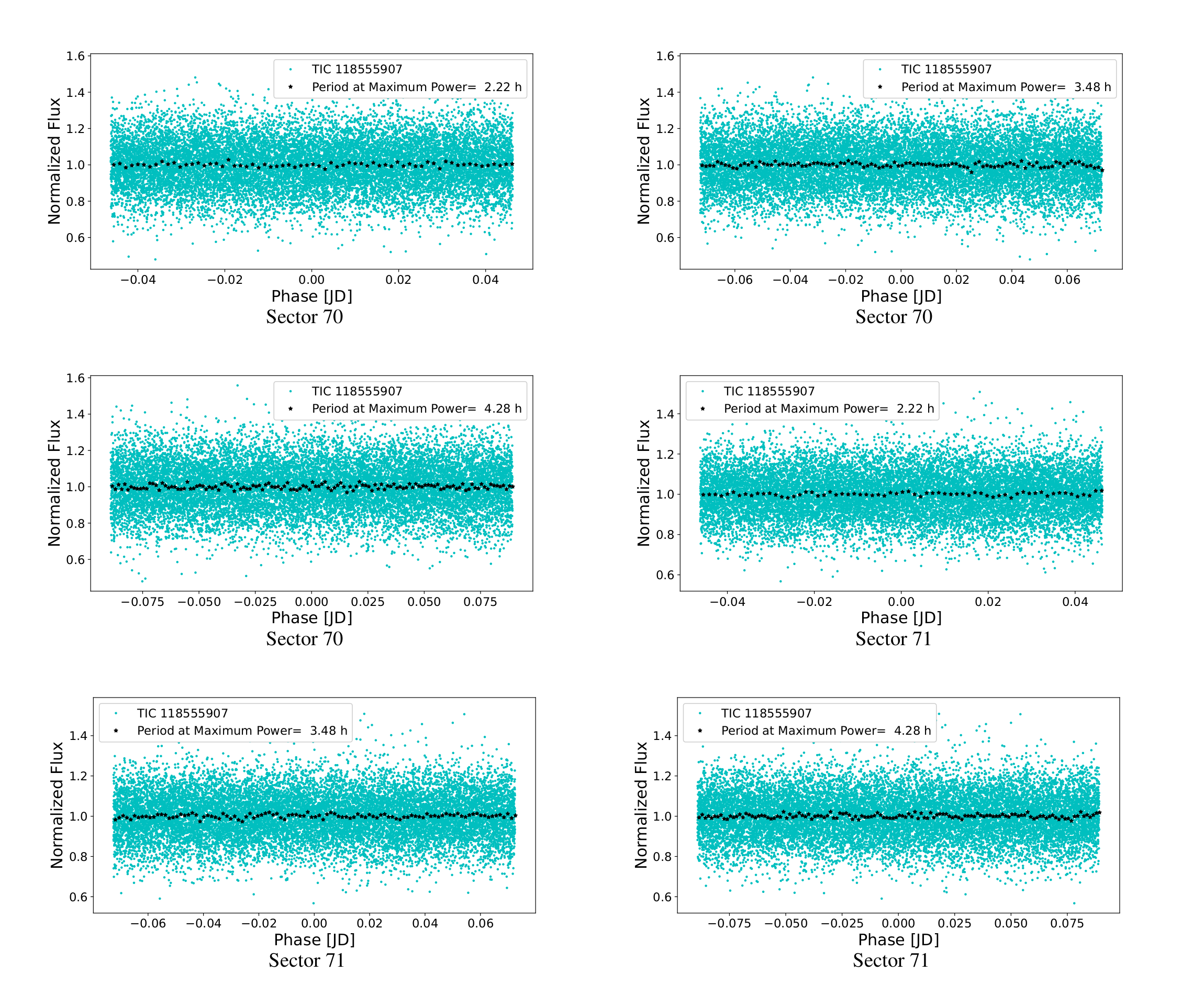}}
    \subcaptionbox{}[.49\linewidth][c]{%
    \includegraphics[width=.98\linewidth]{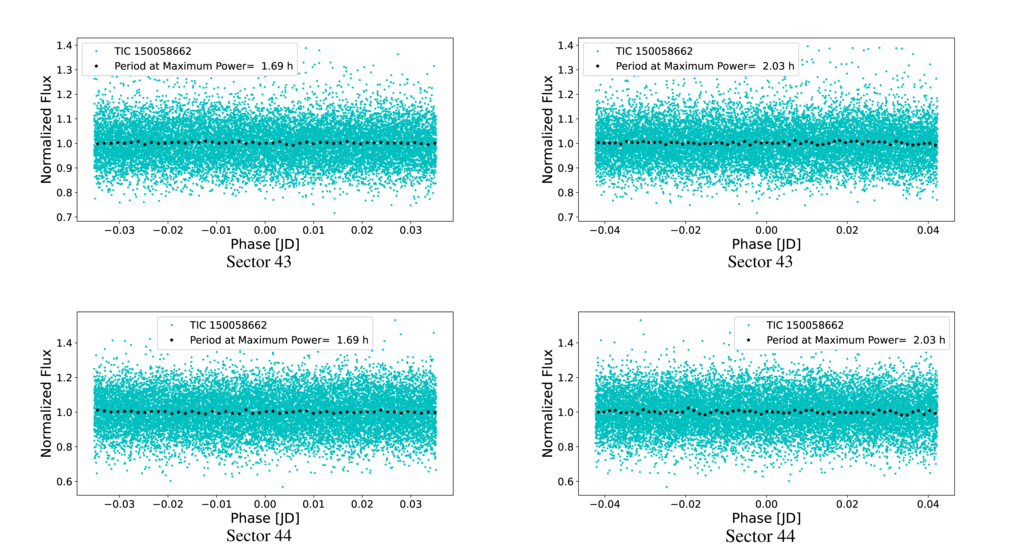}} \quad
   
\end{minipage}
    \caption{(a.) Phase curves for TESS 2-min cadence data for sectors 70 and 71 for CT3, folded with the short hour-scale periods obtained from ground-based data. The turquoise is the TESS normalized fluxes and the black stars are binned data. The period is mentioned inside the plots. Each binned data represents 2 min binning. (b.) The same analysis is done for CT4 with sectors 43 and 44 TESS data. The same colour coding is used here as part `a'.}
   \label{tess_ground_based}
\end{figure*}

\subsubsection{Flares in CFHT-BD-Tau 4 in TESS data}
\label{tess_flare}
\begin{figure*}
\hspace*{-2.0cm}
\begin{minipage}[b]{1.0\linewidth}
  \centering
      \subcaptionbox{}[.45\linewidth][c]{%
    \includegraphics[width=.98\linewidth]{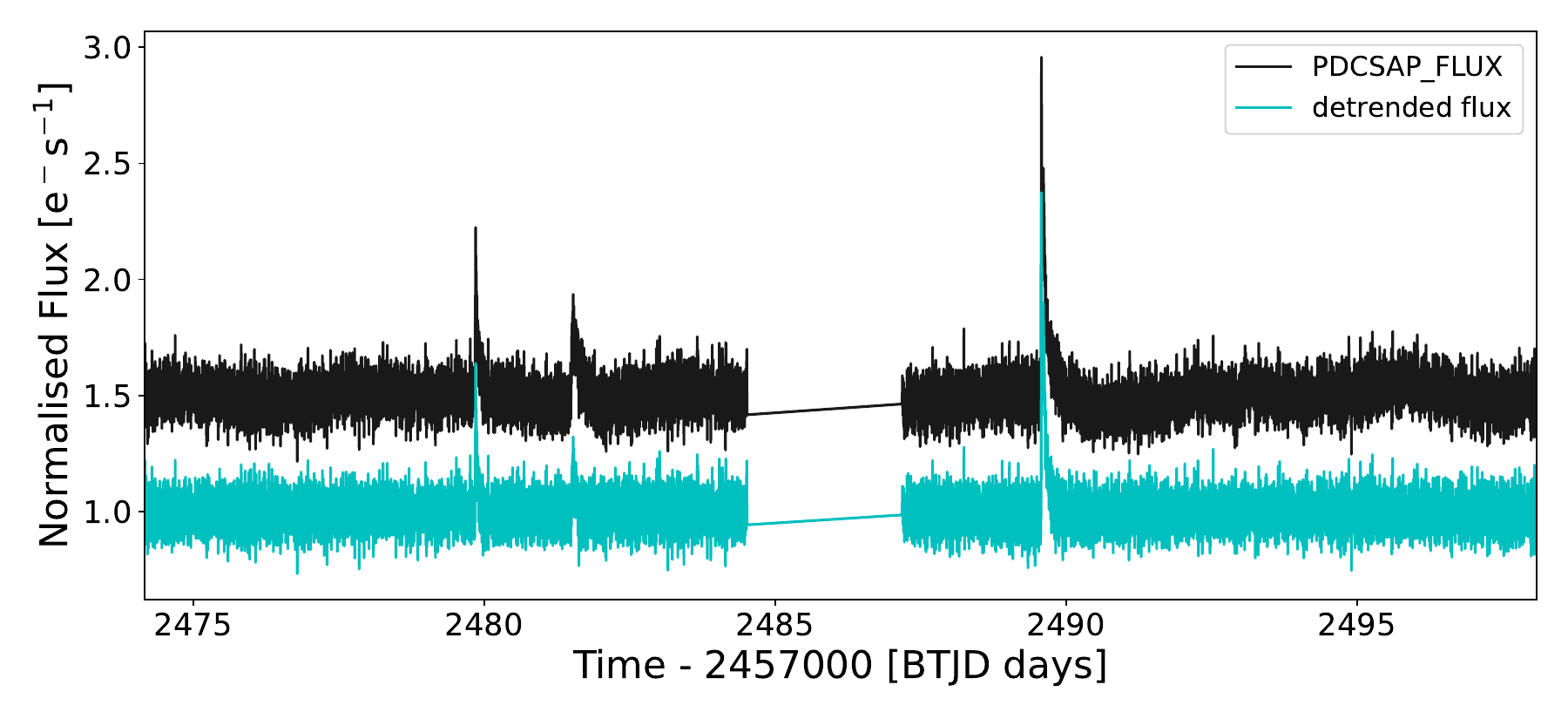}}
      \subcaptionbox{}[.25\linewidth][c]{%
    \includegraphics[width=.98\linewidth]{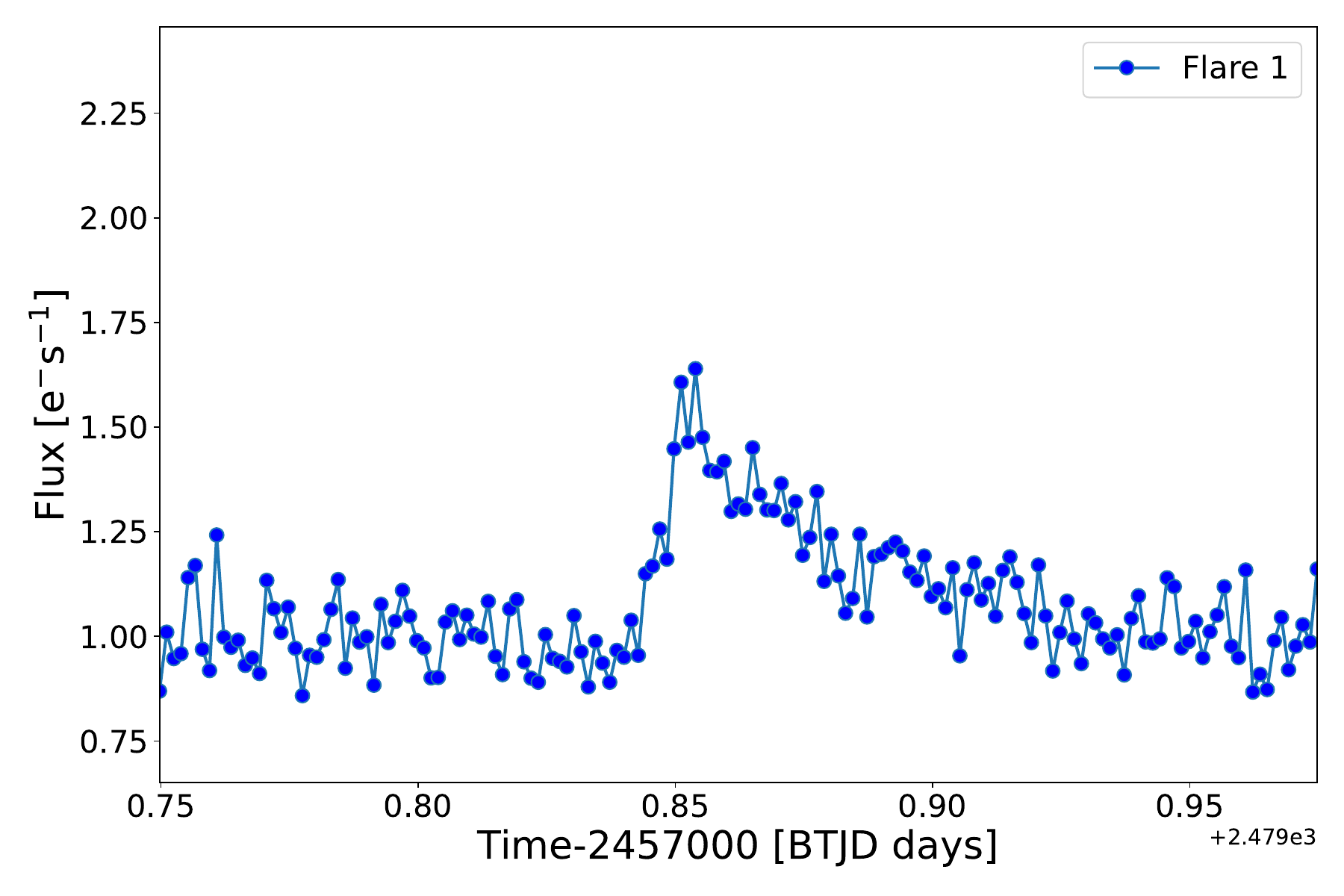}} \quad
	 \subcaptionbox{}[.25\linewidth][c]{%
    \includegraphics[width=.98\linewidth]{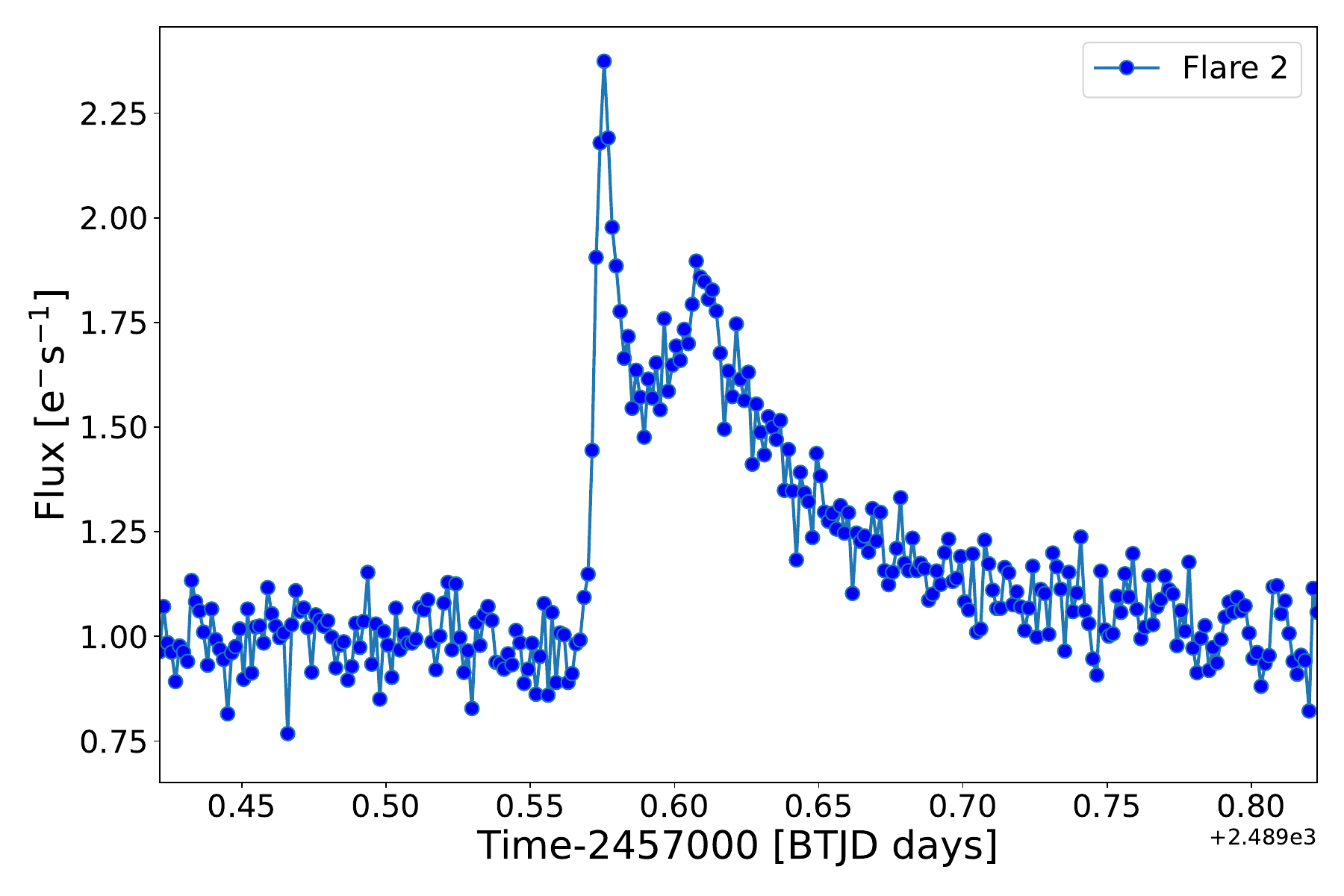}}
      
\end{minipage}
 \caption{Corrected light curves of CFHT-BD-Tau 4 from sector 43 of TIC 150058662 2-min cadence data are shown here (top left). The x-axis is the time in Barycentric TESS Julian Days (BTJD) and the y-axis is the normalized TESS flux ($e^{-}$/s). Black highlights PDCSAP data. Cyan gives the detrended flux ($e^{-}$/s) with no intrinsic variation present. The other two plots show a zoomed-in view of the flares.}	
   \label{flare_corr43}
\end{figure*}

We analyzed the TESS light-curves of CT4 from sector 43 and sector 44 using the {\it altaipony}\footnote{\url{https://altaipony.readthedocs.io/en/latest/tutorials/index.html}} pipeline. We retrieved the PDCSAP light curve (Fig. \ref{flare_corr43}), detrended for instrument systematic. Then the data is detrended for any long-term periodicity. We used the Savitzky-Golay filter ({\it ``savgol'' package}: \citealt{doi:10.1021/ac60214a047}) to the light curve for detrending. The {\it find\_flares()} package detects the flare in the detrended light curve and gives flare duration, starting and ending time of the flare, and equivalent duration for each detected flare mentioned in Table \ref{tess_flare_tbl}. After detrending it, we found two flares in sector 43 data and none in sector 44. The detrended light curve showing the flares is shown in Fig. \ref{flare_corr43}.

\subsubsection{Energy calculation using the Flare as a Blackbody}
\label{energy_calc}
We calculated the energy by integrating the region under the flare, called equivalent duration (ED), which is in the unit of seconds, and multiplied it by the object's luminosity (discussed in \citealp{2021MNRAS.500.5106G}, \citealp{Kumbhakar2023ApJ...955...18K}). ED is the time during which the BD would have emitted the same amount of energy in its quiescent state, as the flare. We got these ED for the flares from {\it altaipony} package as 4883.7s and 851.3s. Assuming that the star is a blackbody radiator, we estimated the flare energy using the flare bolometric luminosity used in \citet{2018ApJ...861...76P}. They calculated the bolometric (UV/optical/IR) energy of the CT4 flare to be $4.9\times10^{30}$ and $5.4\times10^{32}$ erg for an ED of 1s corresponding to $\rm A_v=$ 0.0 and 6.37, respectively. We used them to calculate the flare energy in the order of $\sim 10^{35} - 10^{36}$ erg reported in Table \ref{tess_flare_tbl}. These flare energies can be considered in the super-flare ranges.

To cross-check our flare energy calculation, we followed the well-known method of energy calculation of flares as follows. We calculated the ED by integrating the region under the flare, which is in the unit of seconds, and multiplied it by the object's luminosity. Instead of assuming that the star is a blackbody radiator, we took the synthetic spectra for calculating the luminosity. We estimated the flare energy using the following equations ($1-5$) from \cite{2013ApJS..209....5S} and \cite{2017ApJ...849...36Y}. Following the procedures mentioned in detail in \cite{2021MNRAS.500.5106G} and \cite{Kumbhakar2023ApJ...955...18K}, we calculated the flare energy.
\begin{eqnarray}
L_{\rm flare} &=& \sigma_{\rm SB} T_{\rm flare}^4 A_{\rm flare}~,\\
L'_{\rm star} &=& \int R_{\lambda}F_{\lambda(T_{\rm eff})}d\lambda \cdot \pi R_{\rm star}^2~, \\
L'_{\rm flare} &=& \int R_{\lambda}B_{\lambda(T_{\rm flare})}d\lambda \cdot A_{\rm flare}~, \\
C'_{\rm flare} &=& \frac{ L^{'}_{flare} }{ L^{'}_{star} } = \frac {F_{i}-F_{0}} {F_{0}}~, \\
E_{\rm flare} &=& \int_{\rm flare} L_{\rm flare}(t) dt~.
\end{eqnarray}

where $\sigma_{SB}$ = Stefan-Boltzmann constant, $F_{\lambda(T_{\rm eff})}$ is the flux of BT-Settl synthetic spectra for CT4 ($T_{eff}$ = 2800 K, $logg=$ 3.5), $\lambda$ is the wavelength, $B_{\lambda(T)}$ is the Plank function, $R_{\lambda}$ is the response function of the TESS-band, $L_{\rm flare}$ is  bolometric flare luminosity, $L'_{\rm star}$ is the luminosity of star, $T_{\rm flare}$ is flare temperature, $A_{\rm flare}$ is the area of flare and $C'_{\rm flare}$ is the flare amplitude of the light curve. We took $F_{o}$ as the flux at the average time of the beginning and the end of the flare and $F_{i}$ to be the max amplitude of the flare. The calculated bolometric energy of the CT4 flare is $1.84\times10^{33}$ erg for an ED of 1s which is one order of magnitude higher than previously reported.\\
Since the flare is not a perfect blackbody or the synthetic spectra won't exactly match the star's original spectra, these estimates are not exact and may have an error of a few tens of per cent. 

\begin{table}
\caption{Flare parameters of CFHT-BD-Tau 4 in TESS data}
\label{tess_flare_tbl}
\centering
\begin{tabular}{p{4.5cm}p{2.5cm}p{2.0cm}}\\\hline
Sector & Sector 43 &  Sector 43 \\
Parameters  & Flare 1	 & Flare 2 \\\hline
Flare start [BJD]  & 2489.571444 & 2479.849736  \\
Flare stop [BJD] &  2489.672842 & 2479.874738  \\
E.D. [sec]  &4883.7 $\pm$ 70.1 & 851.3 $\pm$ 37.3  \\
Duration [hour] & 2 h 26 min &  36 min \\
Fractional Amplitude & 1.38  & 0.64 \\
Energy$^a$ [erg] ($A_v= 0.0$) & $2.39\times10^{34}$ & $4.17\times10^{33}$ \\
Energy$^a$ [erg] ($A_v= 6.37$) & $2.64\times10^{36}$ & $4.59\times10^{35}$ \\
Energy$^b$ [erg] ($A_v= 6.37$) & $9.00\times10^{36}$ & $1.7\times10^{36}$ \\
Energy$^c$ [erg] ($A_v= 6.37$) & $3.75\times10^{36}$ & $7.09\times10^{35}$\\
\hline
\end{tabular}
\\$^a$Energy is calculated using the bolometric luminosity from \citealt{2018ApJ...861...76P}. \\
$^{b}$ Energy calculated using the method discussed in section \ref{energy_calc} with radius from \citealt{2018ApJ...861...76P}. A few ten per cent errors are included in energy estimation as discussed in section  \ref{energy_calc}\\
$^c$ Energy is calculated using the same method as $^b$ but with the radius from our SED fitting.
\end{table}

\subsection{Starspot Modeling using \it{TESS} data}
\label{bassman}
\begin{figure*}
\hspace*{-1.cm}
\begin{minipage}[b]{1.\linewidth}
  \centering
      \subcaptionbox{}[.49\linewidth][c]{%
    \includegraphics[width=.98\linewidth]{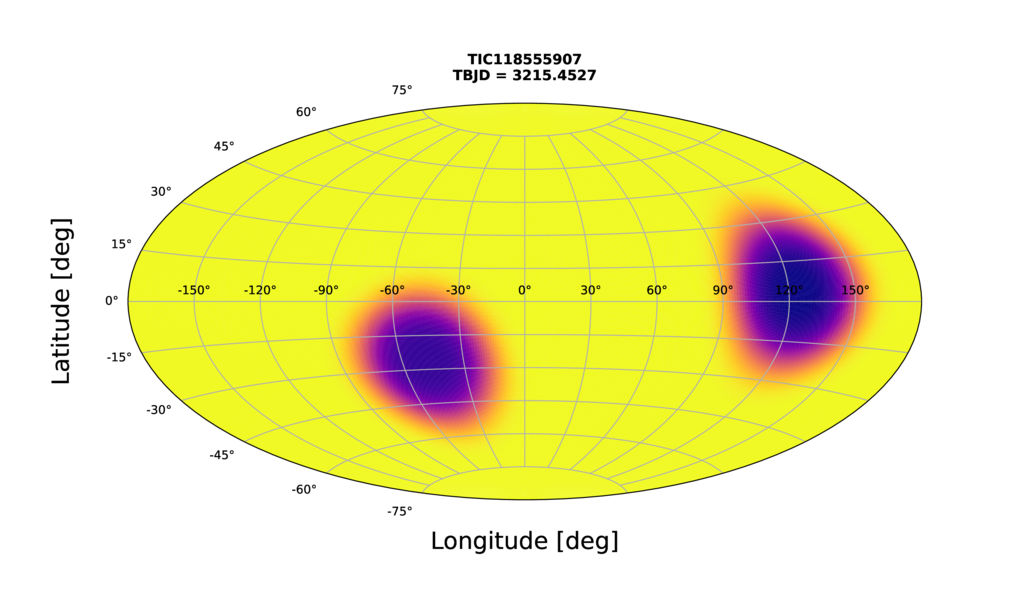}}
      \subcaptionbox{}[.49\linewidth][c]{%
    \includegraphics[width=.98\linewidth]{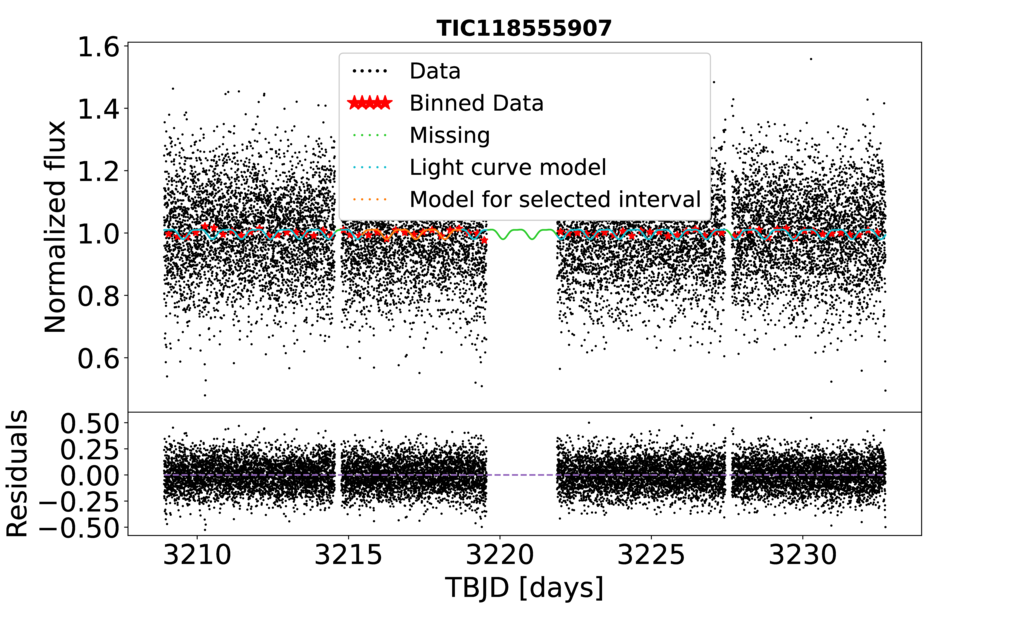}} \quad
	 \subcaptionbox{}[.49\linewidth][c]{%
    \includegraphics[width=.98\linewidth]{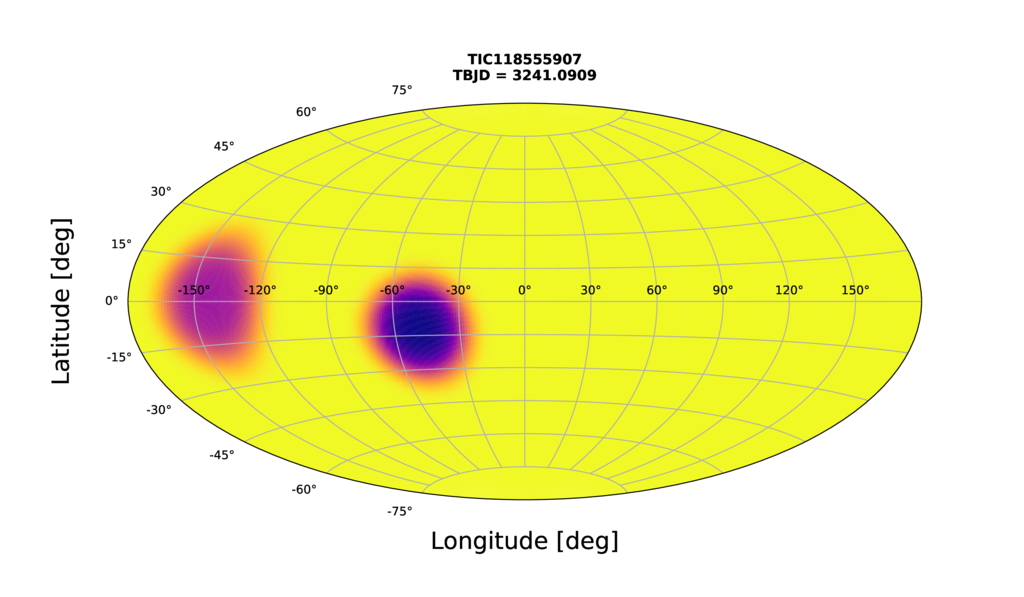}}
    \subcaptionbox{}[.49\linewidth][c]{%
    \includegraphics[width=.98\linewidth]{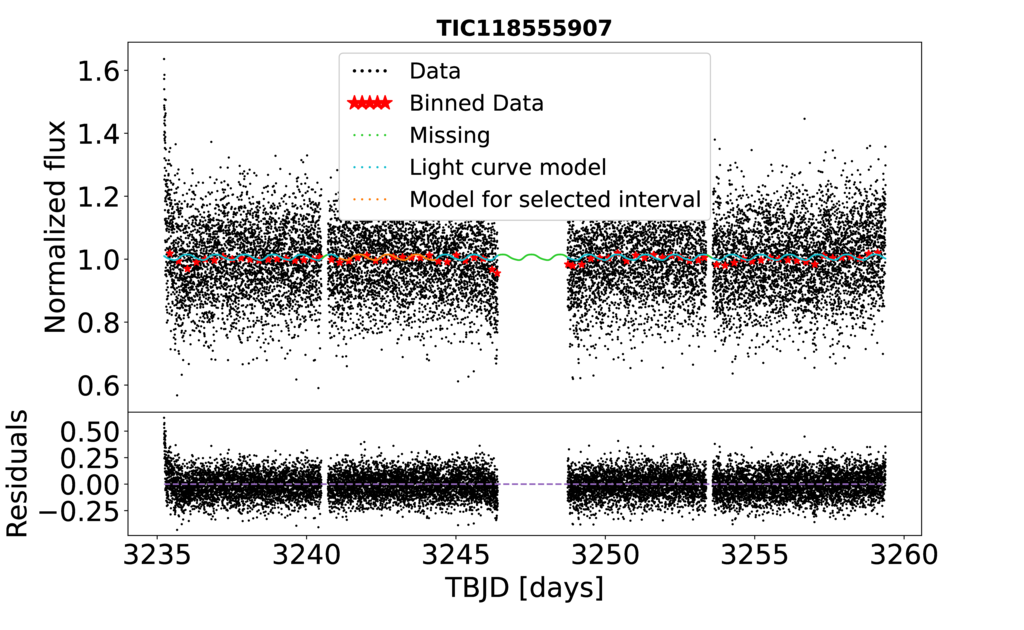}}
      
\end{minipage}
\caption[]{The right panels show the modelled light curves of CT3. The black dots are the TESS data, the red curve represents the recreated light curve, the saffron is the model for the selected interval used in modelling, the cyan is the repetition of the said model for the whole sector data and the green is the model for the gap in TESS data. The left panels present the spots' locations, sizes, and contrasts in Aitoff's projections. The top panels are for Sector 70 and the bottom panels are for Sector 71.}
   \label{spot_ct2}
\end{figure*}

\begin{figure*}
\hspace*{-1.cm}
\begin{minipage}[b]{1.\linewidth}
  \centering
      \subcaptionbox{}[.49\linewidth][c]{%
    \includegraphics[width=.98\linewidth]{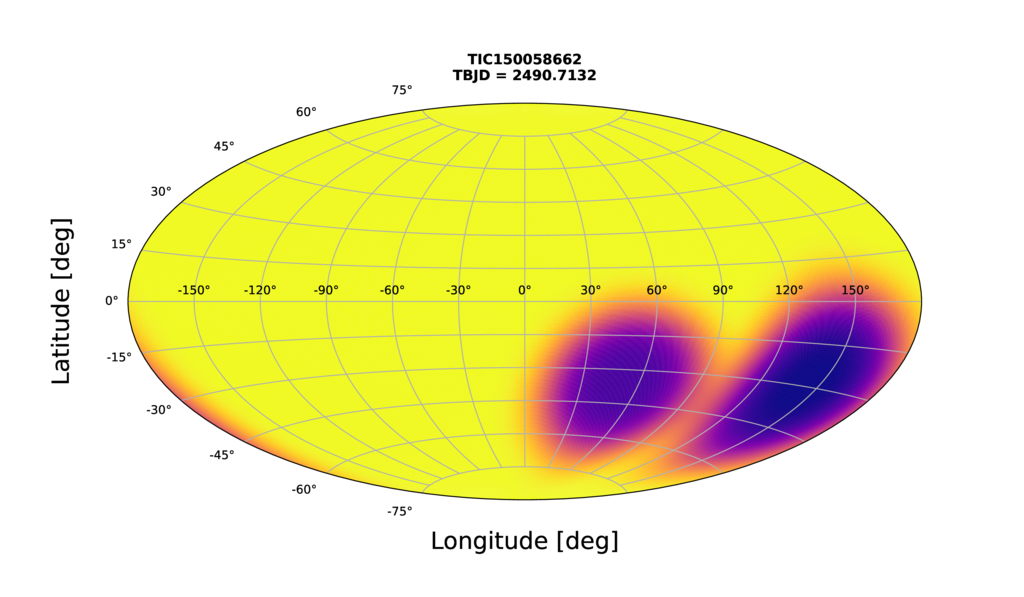}}
      \subcaptionbox{}[.49\linewidth][c]{%
    \includegraphics[width=.98\linewidth]{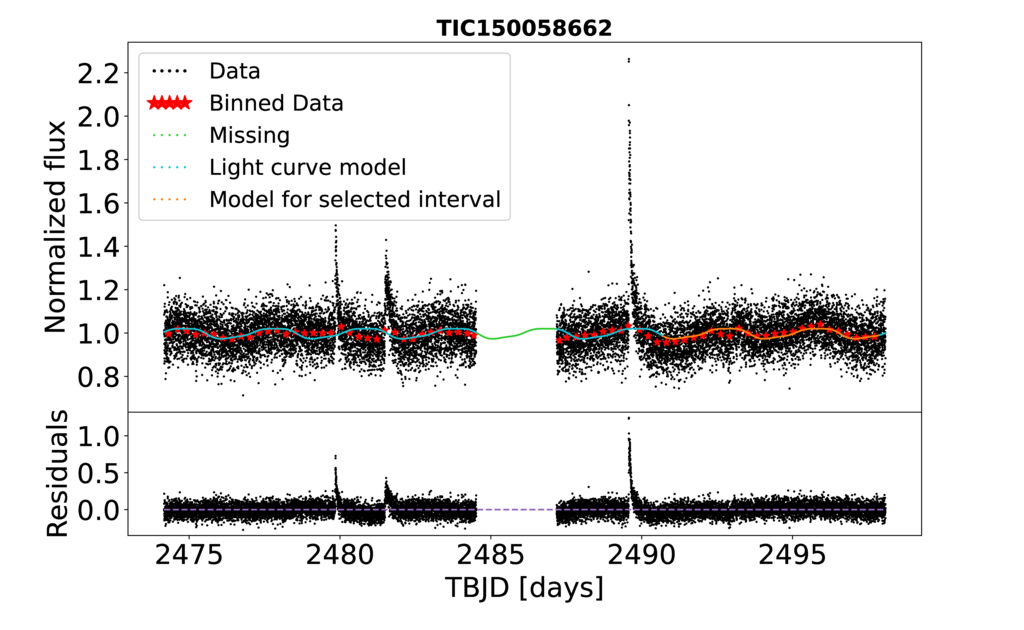}} \quad
	 \subcaptionbox{}[.49\linewidth][c]{%
    \includegraphics[width=.98\linewidth]{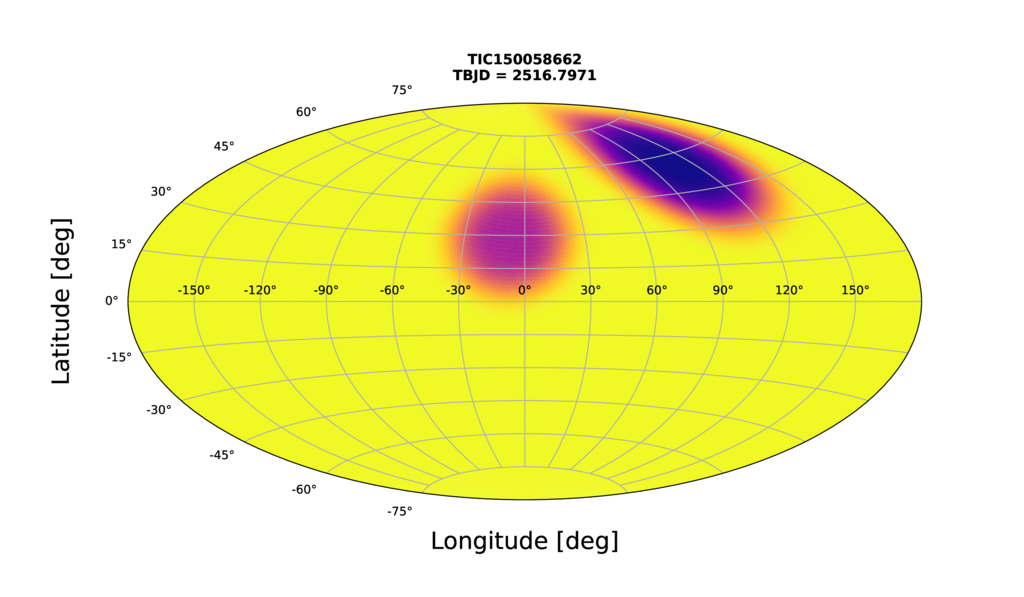}}
    \subcaptionbox{}[.49\linewidth][c]{%
    \includegraphics[width=.98\linewidth]{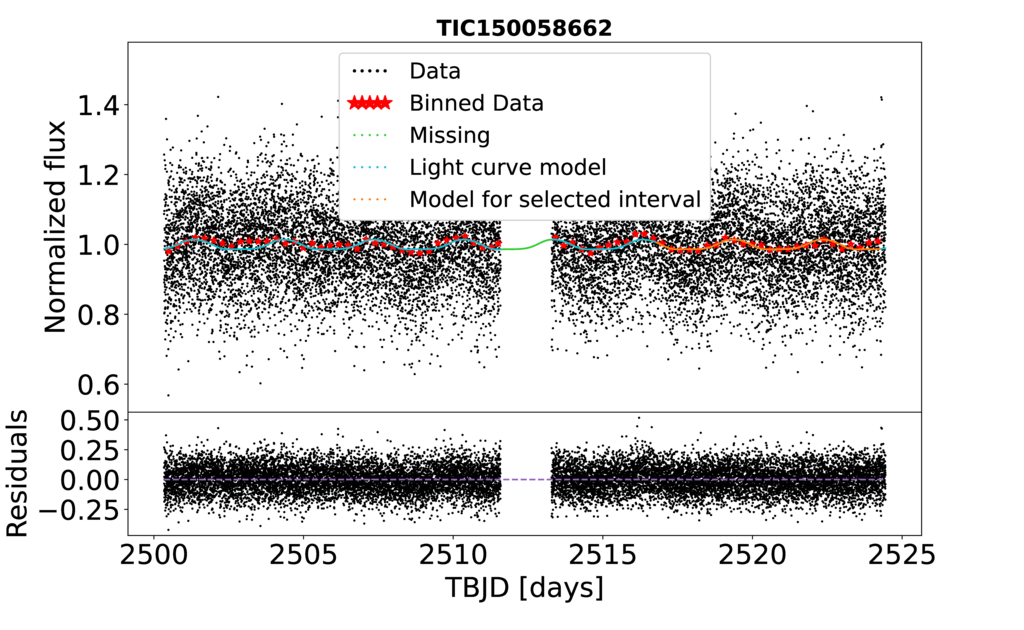}}
      
\end{minipage}
\caption[]{The right panels show the modelled light curves of CT4. The black dots are the TESS data, the red curve represents the recreated light curve, the saffron is the model for the selected interval used in modelling, the cyan is the repetition of the said model for the whole sector data and the green is the model for the gap in TESS data. The left panels present the spots' locations, sizes, and contrasts in Aitoff's projections. The top panels are for Sector 43 and the bottom panels are for Sector 44.}	
   \label{spot1}
\end{figure*}

As mentioned in the introduction, spots corotate with the stellar surface and produce rotational modulation in the light curve with the rotation of the star. Therefore, modelling a star surface can provide, at least theoretically, essential insights into stellar activity and surface dynamics. To map the stellar surface, we have used backward modelling of starspots to recreate the light curves of CT3 and CT4 using {\sc BASSMAN}\footnote{https://github.com/KBicz/BASSMAN} (Best rAndom StarSpots Model cAlculatioN) software \citep{Bicz2022ApJ...935..102B}.  This kind of spot modelling uses time-series data by assuming the spots are static or have differential rotation, their shapes being spherical and discrete. Here, the surface of the stars is expressed as a linear combination of spherical harmonics and the star is described by the vector of spherical harmonic coefficient (for more details. see \citealp{Bicz2022ApJ...935..102B}). We need rotation period, temperature, radius, and inclination angle or rotational velocity  ($v \sin i$) of the star as input parameters in {\sc BASSMAN}. To estimate the inclination angle, we iterated randomly, the value of the period, radius and  $\rm v \sin i$ values within their respective uncertainty using a Monte Carlo simulator \footnote{https://github.com/KBicz/findincmc} and created a probability distribution for the inclination angles. The probability distribution is shown for CT4 in Fig. \ref{spot2}. We fixed the inclination angle (i) as 19 degrees and 67 degrees for CT3 and CT4. By using the Markov chain Monte Carlo (MCMC) method, we derived the starspots' amplitudes, sizes, latitudes, and longitudes. These results help for a better understanding of the surface feature and temporal evolution. Table \ref{table:SED} summarises the results of the spot modelling done with TESS PDCSAP data. The modelled light curve and spot locations are shown in Fig \ref{spot_ct2} and Fig. \ref{spot1}. The model spottedness and mean spot temperature match with the analytical solution from \citep{2019ApJ...876...58N}. For CT3, the spot temperatures are almost constant and total spottedness is changing from $\sim$16.9 \% to 10.0 \% in two sectors. However, we found that the total spottedness varies from $\sim$23\% to 13\% for CT4 from sector~43 to 44, which could be due to the absence of any detected flares in sector 44, while two flares were detected in sector 43. As the magnitude of CT3 and CT4 are near the TESS limiting magnitude, the light curve data is scattered, causing the model to converge at a lower SNR. 
It is interesting to note that, total spottedness reduces from sector 43 to sector 44 of CT4, which also coincides with the fact that flaring happens in sector 43 whereas no flare is detected in sector 44. For CT3, we did not detect any flares and spottedness is lower in both sectors. Therefore it could be inferred from here that, increasing magnetic activities affects flaring, which is reflected in the spot size and structures as well.

Modelling stellar spots is inherently challenging due to degeneracies (\citealp{2013ApJS..205...17W}), meaning different models with varying spot parameters can accurately reproduce a single observed light curve. Therefore, it is crucial for those converting an integrated light curve into a spot distribution to acknowledge and quantify their uncertainties. Although a single model for a star might not reflect the absolute truth, analyzing multiple light curves consistently can uncover trends and relationships that are likely to have physical significance.

\renewcommand{\tabcolsep}{5pt} 
\begin{table*}

\fontsize{8}{8}\selectfont
\caption{Results from spot modelling of \textbf{CT3} and CT4 using BASSMAN}
\label{table:SED}
\hspace{-2.0cm}
\begin{tabular}{p{0.7cm}p{0.7cm}p{0.8cm}p{1.4cm}p{1.4cm}p{1.8cm}p{1.4cm}p{1.2cm}p{1.4cm}p{1.6cm}p{1.6cm}p{0.6cm}p{0.6cm}}\hline

Object &Sector & Spot & Latitude & Longitude &  Temperature (K) & Average Temp (K) & Spot Size (\%)  &Analytical $T_{eff}$ (K) & Total Modelled Spot Size (\%) & Total Analytical Spot Size (\%) & $\chi^2$  & SNR \\\hline

CT3 & 70 & Spot1 & 3.40 & 122.41 &    2426 $\pm$ 448  & 2438 &  8.49 &  2603  & 16.97 & 16.89  &  0.61	& 8.88\\
CT3 & 70	& Spot2 &  -26.13 & -48.33  & 2450 $\pm$ 415  && 8.47   & &&&&\\
CT3 & 71 & Spot1 & 0.19 & -142.81  &  2423 $\pm$ 476  & 2340 & 5.03  & 2603  &10.02 & 8.92  &  0.63 & 10.48  \\
CT3 & 71 &	Spot2 &   -12.02 & -49.31 &  2246 $\pm$ 724  && 4.99 & &&&&\\\hline

CT4 & 43 & Spot1 &  -36.20  &  51.08 &    2690 $\pm$ 183  &2683&  11.83 & 2630 &23.68 &  20.63  &  0.72	& 16.24\\
CT4 & 43	& Spot2 &  -29.14&   143.03  & 2676 $\pm$ 207  && 11.85  & &&&&\\
CT4 & 44 & Spot1 &  55.33 &   106.4  &  2665 $\pm$ 183  &2691& 6.57  & 2630  &  13.09& 13.03  &  0.66 & 10.64  \\
CT4 & 44 &	Spot2 &   28.26&   -7.75  &  2715 $\pm$ 116  && 6.52 &&&&&\\\hline

\end{tabular}

$^1$Total duration of observation. $^{**}$Snapshot observation of CT4 

\end{table*}

\subsection{TESS data with hour-scale periods from ground-based data}
As a test, if the hour-scale periods detected from ground-based data are significant in the high precision TESS data, we generated TESS phase curves folded with I-band hour-scale periods (from Table. \ref{results-table}) in all the sectors (Fig. \ref{tess_ground_based}). As seen from the phase curves, periodic variation is not visible in those. This may be due to the reason that there is no significant hour-scale periodicity present in the data, or due to significant scatter in TESS data, low amplitude hour-scale periodicity is not visible. However, we maintain the inference that the I-band variability is not significant so that the TESS phase curve looks like straight lines.

\section{Discussion}
\label{discussion_sect}
\subsection{Brown Dwarf Environment}
In Table \ref{results-table}, we summarize the periodicities of BDs, detected in our study. It is interesting to note that the periodicities are not constant throughout the observing runs or not even stable throughout our data. For CT2, we detected periodicity on 2 out of 6 nights. On other dates, even though the RMS of the light curves or the light-curve amplitudes ($\sqrt{2}\times RMS$) is high enough, no periodic variation is present in the data. 
We detected 2.99 and 3.01 day periods from TESS 2-min cadence data of CT4 in sectors 43 and 44, respectively, and 0.97 days and 0.96 days for CT3 in sectors 70 and 71 which match very well with previous studies (\citealt{Scholz_2018}, \citealt{2020AJ....159..273R}). These studies also reported 2.93-day periodicity in K2 data of CT3. These day-scale periods are rotational periods of the BDs, which also match their v{\it sin}i values (see section \ref{vsini}). We suggest that evolving surface features might result in this kind of non-repeatable light-curve morphology detected in our ground-based monitoring \citep{2001A&A...367..218B, Clarke2002, 2013ApJ...779..172G} over a stable long rotational period like detected in TESS and K2 data. We did not uncover the day-scale periods by combining our ground-based $I$-band data. This may happen because of significant stochastic variations in a few individual $I$-band light curves (see Table \ref{results-table}), which are unevenly sparse over the years and small sample size compared to the total data span of ten years (Table \ref{log-table}). As discussed in section \ref{periodogram}, the periods detected in our ground-based data could be the result of these stochastic variations over the main signal.

Periodic variations in VLMs are probably caused by asymmetrically distributed magnetically induced cool stable spots co-rotating with the objects, whereas irregular variations or complex variations can be due to strong magnetic fields and accretion hot-spots \citep{1994AJ....108.1906H, refId0}. As these objects are M dwarfs with $T_{eff} \geq 2700$ K, the existence of atmospheric clouds is highly unlikely \citep{2004A&A...421..259S}. This left us with the option of magnetically induced unstable spots to explain the hour-scale variable light-curve morphology. These spots could be slightly hotter or slightly colder than the photosphere. These BDs are capable of sustaining magnetic activity as suggested by previous studies \citep{Delfosse1998A&A...331..581D,2007MmSAI..78..368G}. However, the low amplitude periodic variations (see Table \ref{results-table}) suggest that this might be due to magnetic spot evolution in time scales of days.

Hot spots are unstable because of accretion instability, accretion rate variation, or misalignment of magnetic dipole and rotation axis \citep{refId0}. Irregular or aperiodic variability is common in VLMs where the periodicity detection is marginal and often missing in subsequent observations \citep{Metchev_2015}. Magnetically affected accretion to the stellar surface could give rise to accretion hot spots in BDs and dynamic accretion rates cause erratic light-curve changes \citep{2007M&PS...42.1695Z, 2010MNRAS.409.1307M}. Magnetic activity is common in M dwarfs \citep{Kochukhov2021A&ARv..29....1K}, which is also supported by the flaring of CT4 in our data and previously reported $H_\alpha$ detection \citep{2005ApJ...626..498M} for this BD. Previous studies have also shown that CT4 is found to be accreting, but CT2 and CT3 are non-accretors \citep{2005ApJ...626..498M}. CT4 is hot enough to generate a strong magnetic field to affect the accretion and all three BDS are hot enough to generate a cool or hot magnetic spot to modulate the emitted light \citep{2012ApJ...750..105R} in a timescale shorter than the rotation period. These irregular variations have another effect on period detection. If the star spots are stable but are masked by high-amplitude irregular variation \citep{Kolokolova2015PolarimetryOS}, the light curves would not reflect the low-amplitude periodic variation in the data. Variable extinction from circumstellar disk can also result in either periodic or long timescale variability \citep{2014ApJS..211....3P}. Variable extinction might not always be sinusoidal, unlike variability caused by star spots, but might appear more likely as eclipse-like features \citep{2014ApJS..211....3P}. The rotational speed of the disc determines extinction variations time scales and depending on the radial distance from the star, it could range from a few hours to years \citep{2009MNRAS.398..873S}.\\

Results from our observations suggest that these objects are periodic on some dates and non-periodic on others. Combining these factors with their light-curve amplitude variation from moderate to high on various dates, the possible explanation could be accretion hot-spots in CT4 \citep{refId0} and/or slowly evolving magnetic spots that create such non-repeating light-curve morphology over a stable rotation period of $\sim 3$ days. From spot modelling of the CT4 (Fig. \ref{spot1}), it is seen that the spots are migrated from the southern hemisphere (sector 43 data: top panel) to the northern hemisphere (sector 44: bottom panel). The total spottedness of the surface also changes from 23 $\%$ to 13 $\%$ as the spot size varies but their temperature remains almost constant. This would lead to non-stable variation in the light curve in the short-scale monitoring. For CT3, the spottedness and migration are less compared to CT4 and total spottedness is changing less (than CT4) in sectors 70 and 71 (Fig. \ref{spot_ct2}). The spots are confined to the equatorial region, and as the inclination is very low, equatorial spots have a less significant role to play in flux modulation here.\\

The best-case scenario for the unstable hour-scale signals in CT2 and CT3 could be evolving surface spots that only have low-amplitude modulation over the main signal. Combining this with the fact that the uncertainty in photometry from ground-based monitoring is high enough to mimic this as a periodic signal in short time scales which might be the case for these BDs. We could infer that these short periods might give some insights into variability even at low significance levels. 

\subsection{Colour-magnitude diagram}
\begin{figure*}
\begin{minipage}[t]{0.4\textwidth}
    \centering
    \includegraphics[width=\textwidth, angle=270]{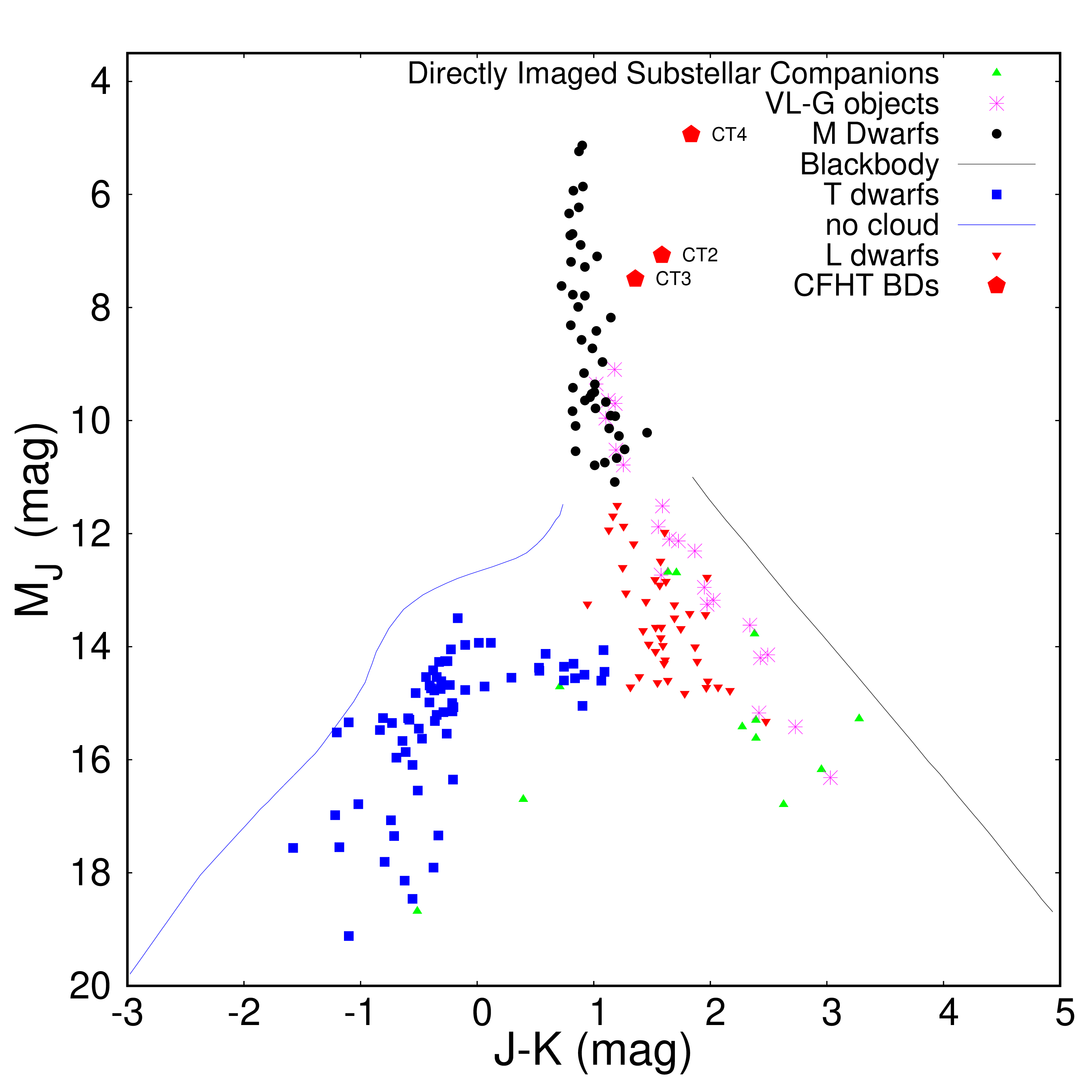}
\end{minipage}
\hfill
\begin{minipage}[t]{0.55\textwidth}
   \caption{Colour magnitude diagram ($J$ - $K$ vs. absolute $J$ magnitude ($M_J$:MKO)) for M, L, and T dwarfs. Data for M, L, and T field dwarfs are from \protect\cite{2012ApJS..201...19D}. This figure is regenerated by using data from figure 1 of \protect\cite{2018ApJ...854..172C} and references therein. Additional data on earlier M dwarfs has been taken from the compilation of \protect\cite{2014PASA...31...43B}. M dwarfs are plotted in black filled `dots', L dwarfs are in red reversed `triangles', T dwarfs as filled blue `squares', low-gravity brown dwarfs as purple `stars', and directly imaged sub-stellar companions as green filled `triangles'. The blue solid line is the computed spectra from Exo-REM assuming no clouds, $log(g) = 4$, with $T_{eff}$ evolving from 400 to 2000 K, and with object radii from the AMES-Dusty evolution model (\protect\citealp{2018ApJ...854..172C}). The `black' solid line was computed assuming a blackbody for the spectrum.} 
    \label{CMD}
\end{minipage}
\end{figure*}

\begin{figure*}
\begin{minipage}[t]{0.6\textwidth}
    \centering
    \includegraphics[width=\textwidth]{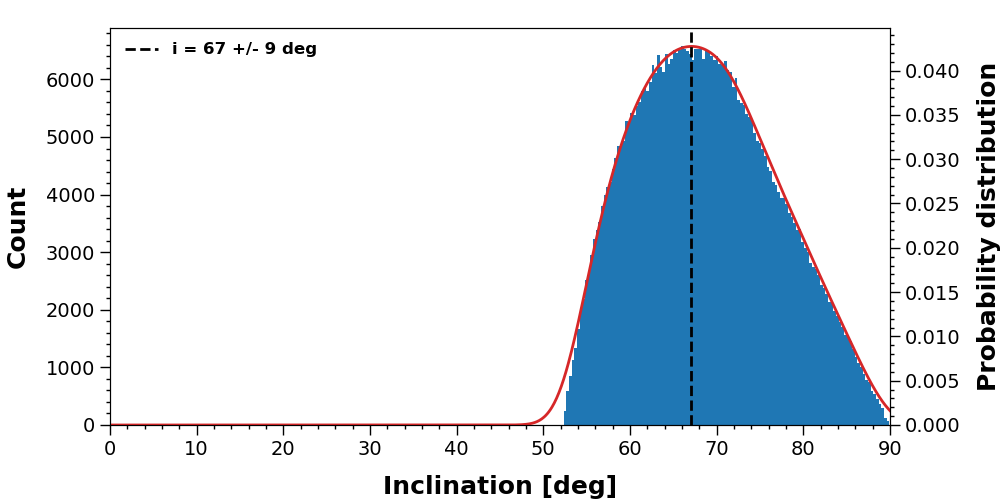}
\end{minipage}
\hfill
\begin{minipage}[t]{0.3\textwidth}
\vspace{-5cm}
\caption{Probability distribution of the inclination angles is shown here for CT4. The distribution is created by iterating the $v \sin i$, period, and radius values randomly within their uncertainty values 1,000,000 times.}
\label{spot2}
\end{minipage}
\end{figure*}

Fig. \ref{CMD} shows the $J$ magnitude as a function of ($J-K$) colour. The three BDs studied here are found to be located at the brighter end of the M-dwarf location, and they are most likely higher mass BDs with larger ($J-K$) colours. A cloud, no-cloud model and blackbody spectrum are displayed in Fig. \ref{CMD} to compare the surrounding environmental conditions of the BDs. \cite{2018ApJ...854..172C} have shown that the existence of clouds of small particles ($\sim$ 3$\mu$m)  grain can re-produce the observed colours of L-T and even in M-type young BDs. The redder colour could be the indicator of a thicker cloud of smaller particles and lower surface gravity than other BDs \cite[see Figure 14 of][]{2018ApJ...854..172C}.

\subsection{Possibility of Magnetic Field in CFHT-BD-Tau 4}
Young BDs are magnetically active, similar to the low-mass stars that result in high-energy flares. In disk-less stars, magnetic reconnection events trigger the flaring processes, while in young M dwarfs or young BDs with disks, the interaction of magnetospheres with disks enables large-scale reconnection events which may produce monstrous flare energies \citep{2018ApJ...861...76P}.
Assuming a similarity between solar flares and flares in VLMS or BDs, we can roughly estimate the magnetic field strength to produce this flare from this equation \citep{2013A&A...549A..66A, 2018ApJ...861...76P},\\
\begin{equation}
E_{\rm flare} = 0.5 \times 10^{32} \left( \frac{B_z^{max}}{1000G} \right)^2  \left( \frac{L^{bipole}}{50Mm} \right)^3   \\
\end{equation}

where $ E_{ flare} $ is the bolometric flare energy, and $\rm L^{bipole}$ is the linear separation between bipoles. If we take $\rm L^{bipole}$ to be equal to $\pi R$ where $ R = 0.65 R_{\odot}$ (from Table \ref{SED_table}), as the maximum distance between a pair of magnetic poles on the surface of CT4, a magnetic field of $\sim$ 1.66 kG is required to produce the flare with energy, $E_{\rm flare}  =  2.64\times10^{36} ~erg$. If we used the corresponding parameter values from this work, using $ R = 0.42 R_{\odot}$ as the maximum distance of a pair of magnetic poles, then a magnetic field of $\sim$ 3.39 kG is required to produce the flare with energy, $E_{\rm flare}  =  3.57\times10^{36} ~erg$. Previously \cite{2018ApJ...861...76P} reported the magnetic field to be 13.5 kG with a flare energy $\sim 2.1 \times 10^{38} ~erg$ with a radius $ R = 0.65 R_{\odot}$.\\ 

Outbursts from magnetospheric accretion could mimic a flare-like event. Following  \cite{Gullbring_1998}, the rate of accretion can be estimated for a star-disk system using this equation,
\begin{equation}
L_{\rm flare} = \frac{GM_{*}\dot{M}}{R_{*}} \left(1 - \frac{R_{*}}{R_{in}} \right)   \\
\end{equation}
where $M_{*}$ is the mass of the BD, $L_{\rm flare}$ is flare luminosity, $R_{*}$ is the radius of the BD, $R_{in}$ is the inner radius of the accretion disk, $\dot{M}$ is the accretion rate and $G$ is the gravitational constant. Using Equation (4) of section \ref{energy_calc}, we estimated $L_{\rm flare}$ for the highest flare energy, $E_{\rm flare}  = 3.16\times10^{36} ~erg$, detected in TESS data. We have taken  $R_{in}=2.5 R_{*}$ as suggested by \cite{2018ApJ...861...76P}. \cite{1991ApJ...370L..39K} provided the classic relation of how the magnetic field is related to stellar mass and radius, accretion rate, and angular velocity (Equation 3 in the paper). This relation is also connected to the inner radius of the disk, i.e., the radius where the disk is truncated. Using representative values for a model T Tauri Star, \cite{1991ApJ...370L..39K} predicts a magnetic field of kilogauss strength on the surface of the star. The inner edge of the disk is then at a radius of $R_{in} = 2 R_{\ast}$. Changing this $R_{in}$ would mean that the magnetic field strength is either higher or lower. We do not have an accurate estimation of the inner radius of the disk of CT4. \cite{2003ApJ...585..372L} suggested $R_{in} \sim (2 - 3) R_{\ast}$ for young brown dwarfs. Their estimate is cool objects in IC 348 and Taurus that have spectral types M6–M9 and ages 5 Myr. So we have taken $2.5R_{\ast}$ for CT4. Using these values, we estimated the accretion rate $\log (\dot{M})$ = -6.64 $M_{\odot}~yr^{-1}$ for  $R_{*}$= $0.65R_{\odot}$ and $\log (\dot{M})$ = -6.27 $M_{\odot}~yr^{-1}$ for $R_{*}$= $1.51R_{\odot}$. If we take the radius as 0.42 $R_{\odot}$, mass as 0.066$M_{\odot}$ (from SED fitting in this work, see Table. \ref{SED_table}) and use the $\rm L_{flare}~(=E_{flare}/ED)$ from last column of Table. \ref{tess_flare_tbl}, then the estimated accretion rate is $\log (\dot{M})$ = -7.69~$M_{\odot}~yr^{-1}$. This accretion rate is in accordance with previously reported value $\log (\dot{M})$ = -6.70 $M_{\odot}~yr^{-1}$ by \cite{2018ApJ...861...76P}.

Flares in CT4 or, more generally, in VLMs and BDs are of increasing importance in the field of the dynamical and chemical evolution of the circumstellar disks around them. Flares cause enhanced UV and X-ray radiation, which is similarly important in the field of habitability of planets around the most numerous and planet-hosting M dwarf stars and BDs.

\subsection{Rotation, v$sin$i and Orbital Inclination} 
\label{vsini}

CT2, CT3 and CT4 have spectroscopic v$sin$i values of 8, 12, and 11 km/s, respectively, with an error of $\pm2$ km/s \citep{2005ApJ...626..498M}. Adopting radii of 0.47 $R_{\odot}$ for CT2 (\citealp{2014A&A...570A..29B}) and 0.42 $R_{\odot}$ for CT3 (\citealp{2014A&A...570A..29B}) and 0.65 $R_{\odot}$ for CT4  (radii from \citealp{2014A&A...570A..29B}), and the rotation periods of 3.01 days, 0.96 days and 2.9 days for CT2, CT3 and CT4 respectively (this work, \citealt{Scholz_2018}, \citealt{2020AJ....159..273R}), we calculated the inclination angle of CT3 and CT4 as 67 $\pm$ 9 degrees and 79 $\pm$ 5 or viewed almost edge-on. For CT2, the inclination angle was estimated as $\sim$ 19 $\pm$ 2 degrees. This estimation includes a few ten per cent errors in the parameters used.

\subsection{Deuterium Burning Pulsation}
The short hour-scale variability observed in the ground-based short-baseline monitoring of the three BDs does not match the day-scale rotational periods found with TESS and K2 data. VLMs and BDs are known to rotate on the hour-to-day scale period and previously probed for pulsation \citealp{2014A&A...566A.130C, 2014ApJ...796..129C}. One way, this discrepancy can be addressed is if the deuterium burning instability \citep{2005A&A...432L..57P} or $\epsilon$-mechanism \citep{1964AnAp...27..141G} is taken into consideration. In the mass range of 0.2 to 2.0 $M_{\odot}$, PMS stars could become pulsationally unstable during deuterium burning, affecting the radiation and making them variable \citep{1972A&A....19...76T, 2005A&A...432L..57P}. This instability is induced in the core of fully convective VLMs and BDs due to the high sensitivity of nuclear energy generation rate to temperature ($\sim T^{12}$). A small temperature perturbation could induce an order of magnitude higher energy variations. \citet{2005A&A...432L..57P} found from their stability analysis that in the mass range of BDs, a pulsation period varies between $\sim$1h to $\sim$5h, which in turn could be observed as a short periodicity in the data with observable amplitude. 
From this discussion, if we assign the day-scale periods of CT2, CT3 and CT4 as rotational periods or due to the presence of surface spots, then deuterium burning instability could lead to unstable hour-scale periods.

\section{Summary}
\label{summary}
In this paper, we presented a time-series photometric study of three known young M-type BDs: CT2, CT3 and CT4 in the Taurus star-forming region. We summarized the results as follows:\\

 From the $I$-band light-curve analysis of the three BDs, we detected hour-scale photometric variability using the Lomb-Scargle periodogram method. For each object, we find both periodic and non-periodic light-curve morphology in various nights of observation. We find that the detected periods are varied in various observing dates ranging from 1.5 to 4.8 hours. From our Monte Carlo simulation of the light curves, we find these periods might give some insights into variability even at low significance levels. However, the reader should use these period parameters with caution.\\
 
 In this work, TESS data shows a day-scale stable period of 2.99 and 3.01 days for CT4 in sectors 43 and 44, and 0.97 days and 0.96 days for CT3 in sectors 70 and 71, respectively, which is also reported in previous studies. CT2 showed day scale periodicities of 2.93 days in previous studies. These factors, along with their light-curve morphology from hour-scale monitoring, suggest that there might be accretion hot spots (for CT4) co-rotating with the objects that create such non-repeating light-curve morphology over the stable day-scale rotation periods. Another possible reason for the unstable hour-scale periods (at least a few of them) could be the deuterium burning instability which can also introduce pulsation periods of $\sim$1 - 5 h with detectable amplitude. \\
 
 Periodic variability in low-mass objects is due to rotational modulation of incoming flux by asymmetrically distributed star spots on the surface of the objects. The aperiodic variability is thought to be caused by accretion from a circumstellar disk and/or evolving magnetic star spots. These accretions cause non-periodic variations and flaring that are commonly found in BDs. From our spot modelling of CT4, the changing spottedness and location of the spot might be responsible for the stochastic changes in the I-band data as this changes the light curve morphology. For CT3, spot modelling shows spot location and sizes are almost constant or changing less in TESS sector 70 and 71.\\
 
 In $M_J ~vs.~(J-K)$ colour-magnitude diagram, these BDs are located at a brighter end of the M-dwarf region, suggesting the existence of a cloudy atmosphere of small particles ($\sim3 \mu ~m$).\\
 
 From TESS 2-min cadence data we detected two super-flare events with energies $7.09\times10^{35}$ erg and $3.75\times10^{36}$ erg. We approximately estimated the magnetic field as $\sim3.39 ~kG$, needed to generate such energetic ($\sim 10^{36}$ erg) flare. If outbursts from magnetospheric accretions from disks result in such energetic flares like the one we detected, then we estimate an accretion rate of $\log (\dot{M})$ = -7.69 $M_{\odot}~yr^{-1}$.\\
 
  Total spottedness reduces from sector 43 to sector 44, which also coincides with the fact that flaring happens in sector 43 whereas no flare is detected in sector 44. For CT3, we did not detect any flares and the total spottedness is lower than CT4.

\section*{Acknowledgements}
This research work is supported by the S N Bose National Centre for Basic Sciences under the Department of Science and Technology, Govt. of India. The authors are thankful to the Joint Time Allocation Committee (JTAC) members and the staff of the 1m-ST and 1.3-m Devasthal optical telescope operated by the Aryabhatta Research Institute of Observational Sciences (ARIES, Nainital), the VBO Time Allocation Committee (VTAC) members and the staffs of the 1.3-m J.C. Bhattacharyya Telescope, the HCT Time Allocation Committee (HTAC) members and the staff of the Himalayan Chandra Telescope (HCT), operated by the Indian Institute of Astrophysics (IIA, Bangalore). SG is grateful to the Department of Science and Technology (DST), Govt. of India for their Innovation in Science Pursuit for Inspired Research (INSPIRE) Fellowship scheme. SG is grateful to Dr. Anuvab Banerjee for the discussions on the statistical analysis of light curves. This publication makes use of VOSA, developed under the Spanish Virtual Observatory project supported by the Spanish MINECO through grant AyA2017-84089. VOSA has been partially updated by using funding from the European Union's Horizon 2020 Research and Innovation Programme under Grant Agreement nº 776403 (EXOPLANETS-A). This paper includes data collected with the TESS mission, obtained from the Mikulski Archive for Space Telescopes (MAST) data archive at the Space Telescope Science Institute (STScI), which is operated by the Association of Universities for Research in Astronomy, Inc., under NASA contract NAS 5–26555. This work has made use of data from the European Space Agency (ESA) mission {\it Gaia} (\url{https://www.cosmos.esa.int/gaia}), processed by the {\it Gaia} Data Processing and Analysis Consortium (DPAC, \url{https://www.cosmos.esa.int/web/gaia/dpac/consortium}). Funding for the DPAC has been provided by national institutions, in particular the institutions participating in the {\it Gaia} Multilateral Agreement.

\section*{Data Availability}
Data were obtained using Indian telescope facilities: the 1-m ST, ARIES, Nainital; 1.3-m DFOT Devasthal, Nainital; 1.3-m JCBT, Kavalur and 2-m HCT, Hanle, Ladakh. Data are not publicly available but will be provided upon request. This paper includes publicly available data collected with the TESS mission, obtained from the Mikulski Archive for Space Telescopes (MAST) data archive at the Space Telescope Science Institute (STScI).

All of the TESS data presented in this article were obtained from the Mikulski Archive for Space Telescopes (MAST) at the Space Telescope Science Institute. The specific observations analyzed can be accessed via \href{http://dx.doi.org/10.17909/cp03-6167}{this link (click here)}.


\vspace{5mm}




\appendix
\section{Non-variable light curves}
\label{non_var_LC}
This section shows the non-variable light curves observed in the optical $I$ band (Fig. \ref{non_var_LC}). We classified these light curves as non-variable as their periodogram analysis shows no `significant' peaks after correcting for aliases.
\begin{figure*}
\begin{minipage}[b]{1\linewidth}

  \centering
      \subcaptionbox{CT2: 1.3-m DFOT, 08.12.2018}[.3\linewidth][c]{%
    \includegraphics[width=.98\linewidth]{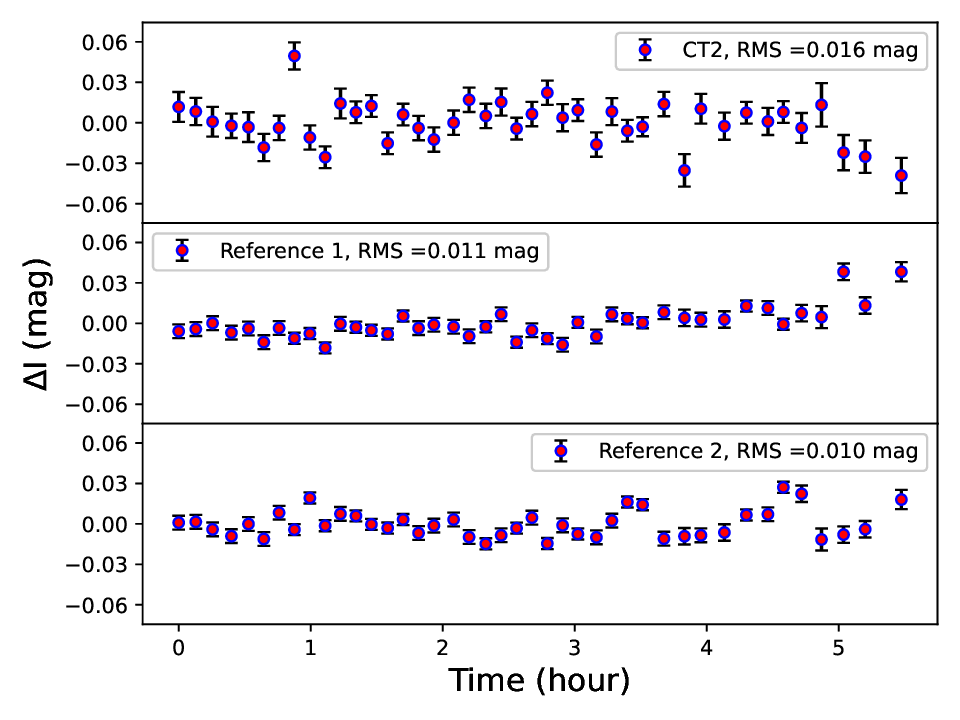}}
      \subcaptionbox{CT2: 1.3-m JCBT, 01.01.2019}[.3\linewidth][c]{%
    \includegraphics[width=.98\linewidth]{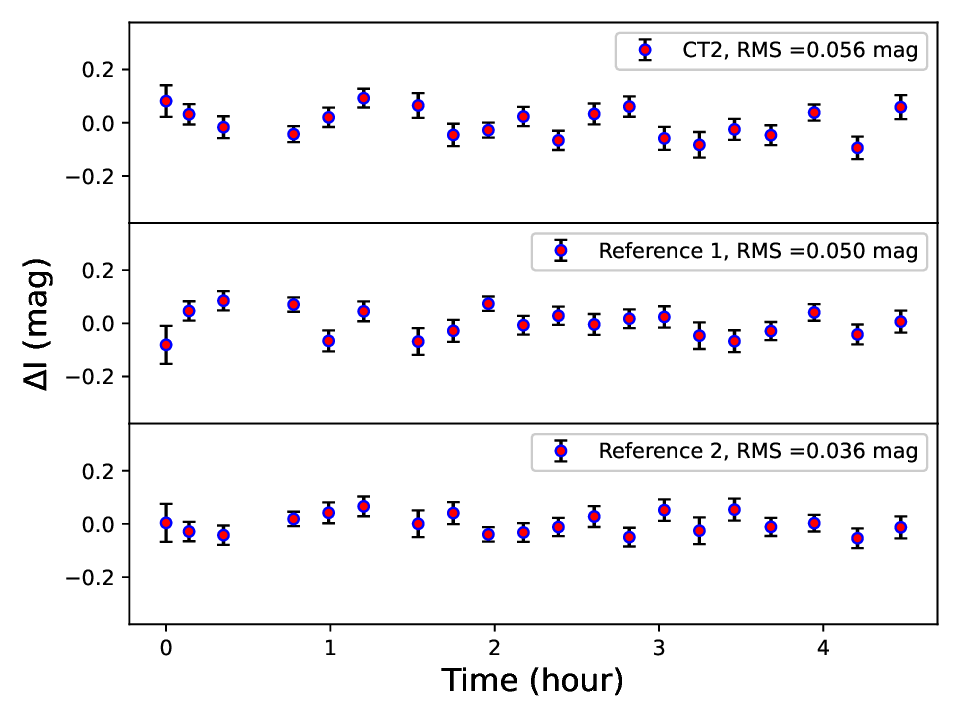}} 
       \subcaptionbox{CT2: 1.3-m JCBT, 02.01.2019}[.3\linewidth][c]{%
    \includegraphics[width=.98\linewidth]{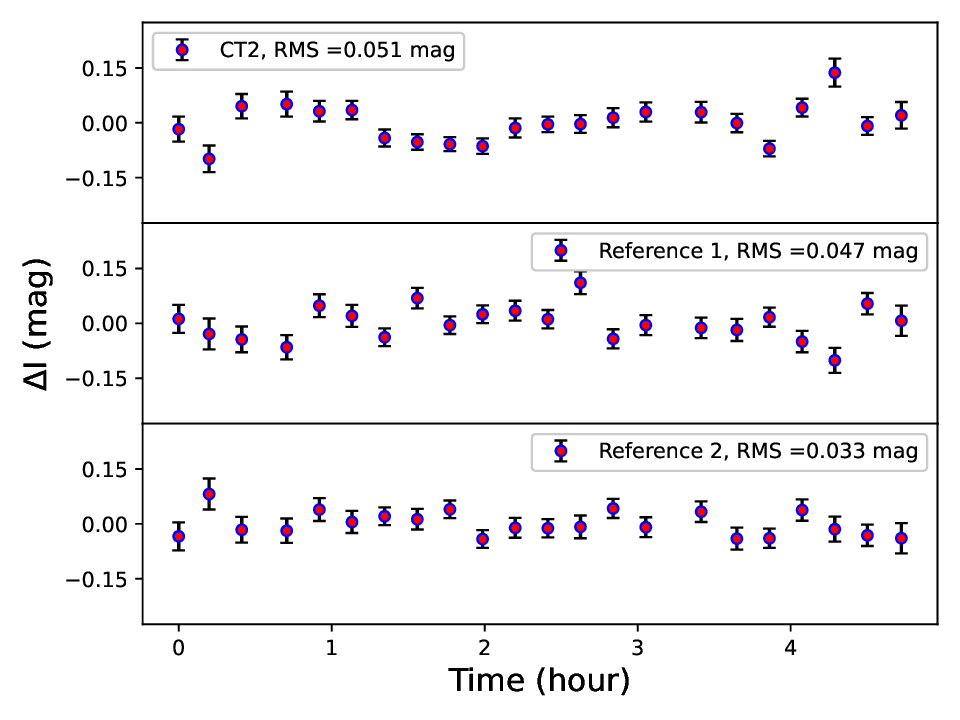}} 

      \subcaptionbox{CT3: 1.3-m DFOT, 08.12.2018}[.3\linewidth][c]{%
    \includegraphics[width=.98\linewidth]{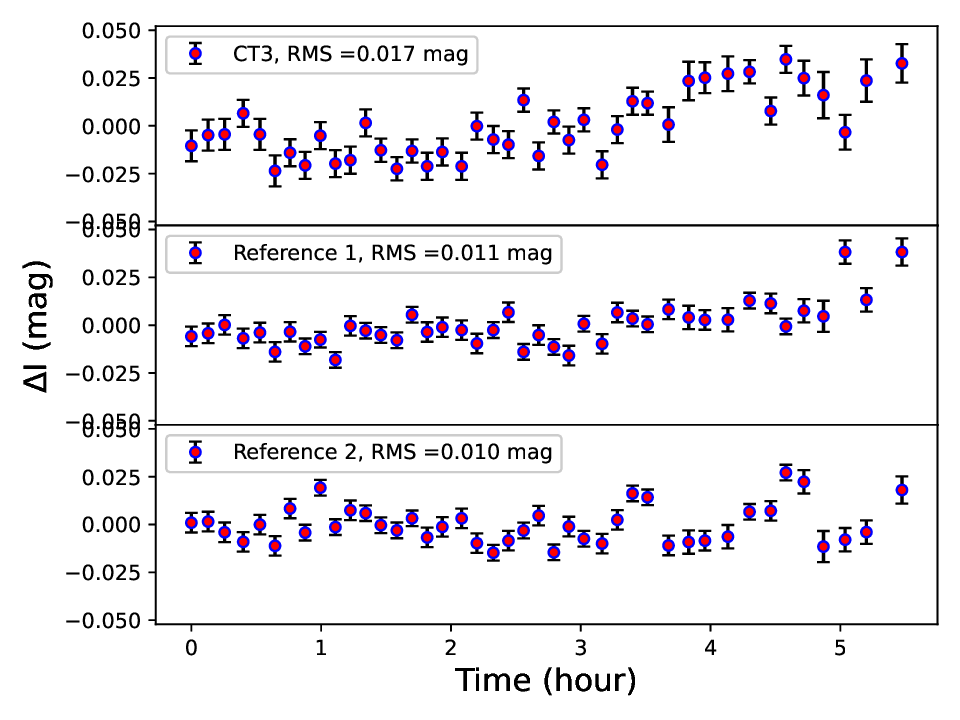}}
      \subcaptionbox{CT3: 1.3-m JCBT, 01.01.2019}[.3\linewidth][c]{%
    \includegraphics[width=.98\linewidth]{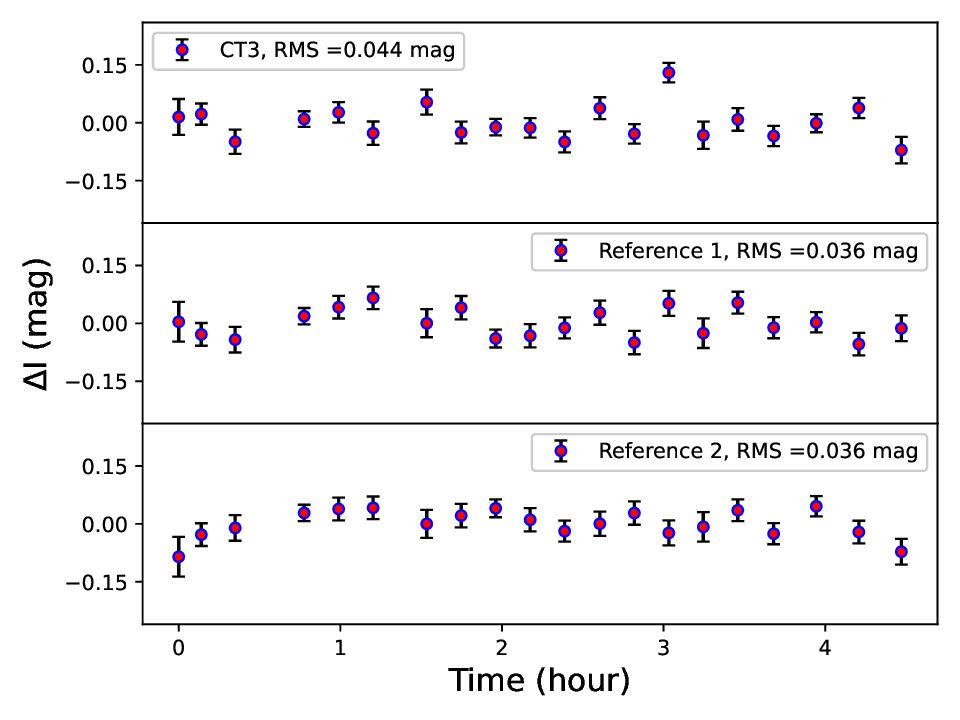}} 
       \subcaptionbox{CT3: 1.3-m JCBT, 02.01.2019}[.3\linewidth][c]{%
    \includegraphics[width=.98\linewidth]{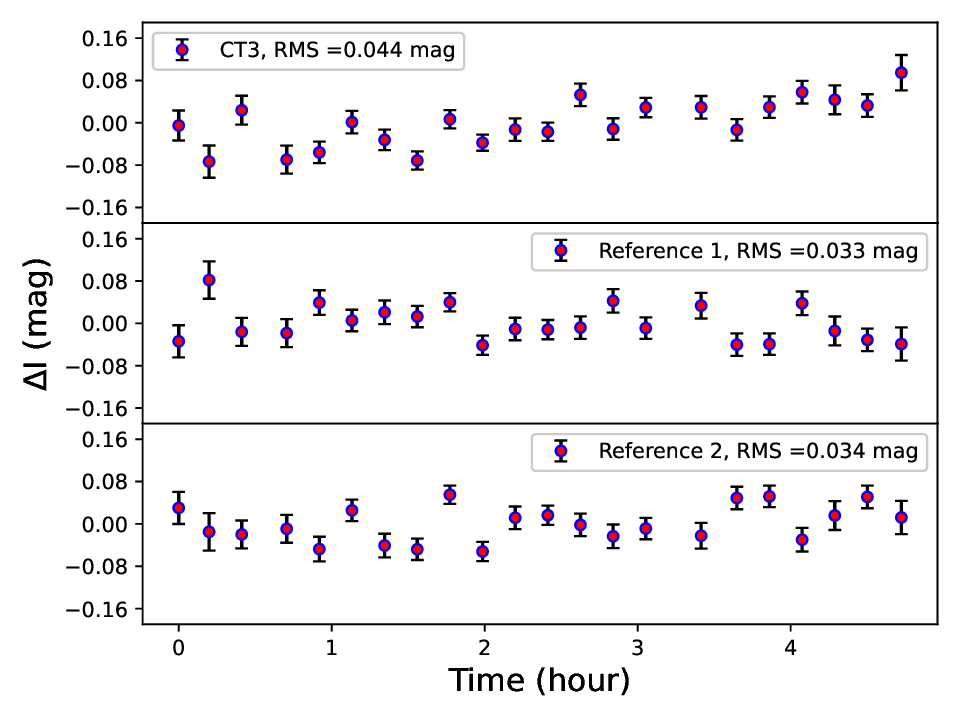}} 

          \subcaptionbox{CT4: 1.3-m DFOT, 09.12.2018}[.3\linewidth][c]{%
    \includegraphics[width=.98\linewidth]{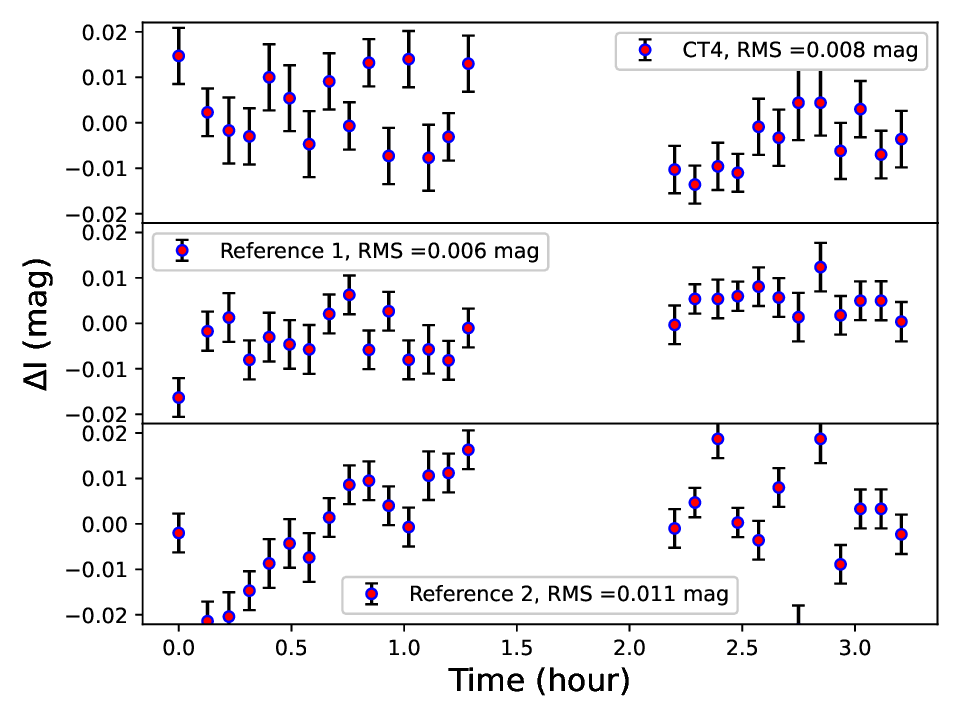}}

\end{minipage}
 \caption{Non-variable light curves of the three brown dwarfs are shown. Two non-variable reference light curves are shown in the bottom panels of each light curve. The source name, dates and telescopes used to obtain the data, are mentioned underneath each panel.}	
   \label{non_var_LC_fig}
\end{figure*}


\bibliography{Arxiv}{}
\bibliographystyle{aasjournal}

\section{Supplementary Material}
\label{supl1}
\section*{Results of periodogram analysis using Monte Carlo Simulation}
All the figures were generated from the same analysis already discussed in section \ref{periodogram}. We show here the same analysis done on the sources on other dates of I-band observations.

\begin{figure}
    
\begin{minipage}[b]{.7\linewidth}

  \centering
    \includegraphics[width=.9\linewidth]{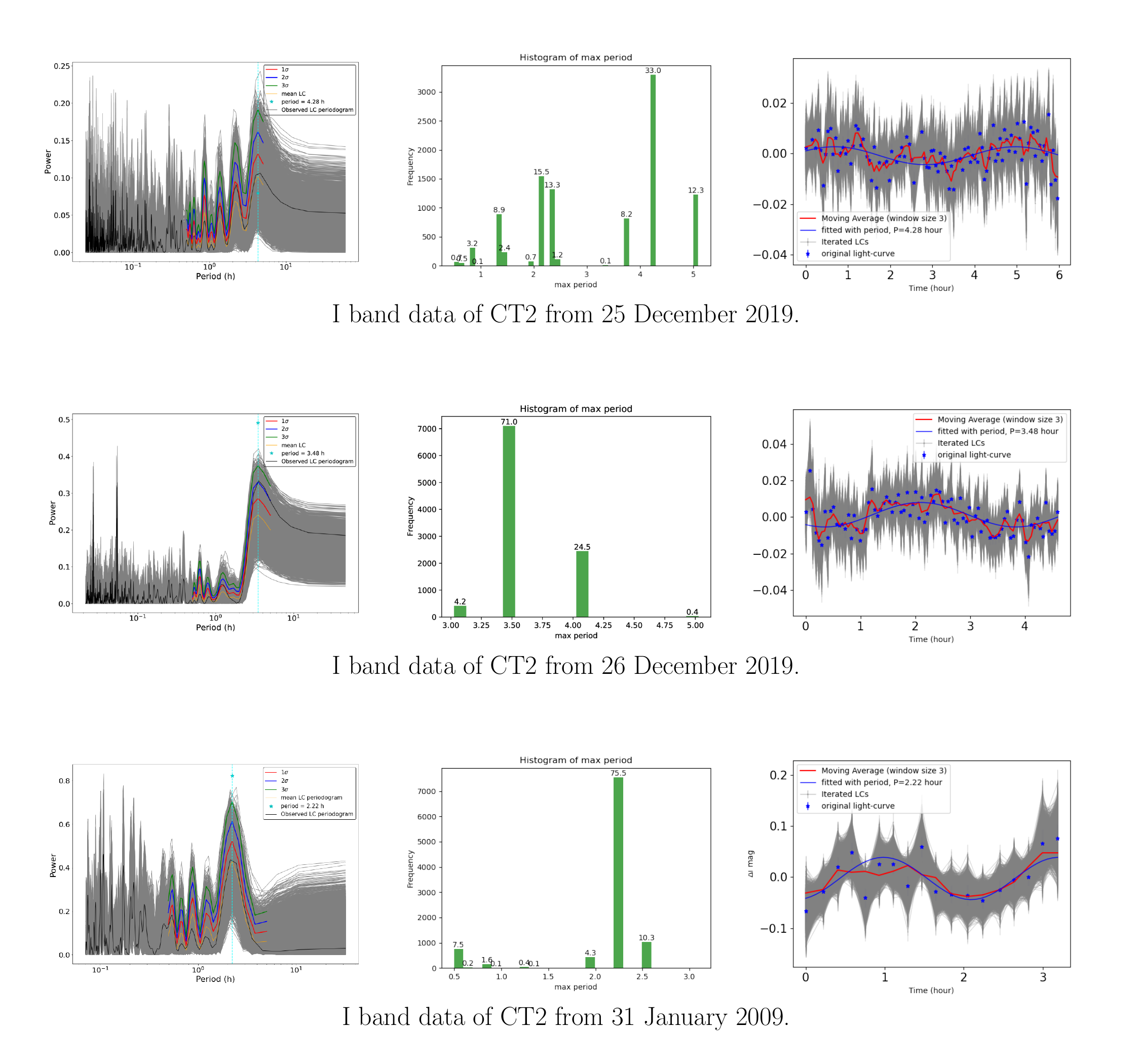}

\end{minipage}
 \caption{ The top right panel represents the 10000 iterated LCs, shown in grey with blue stars as the original LCs of CT2 on various dates. The red line represents a moving average with a window size of 3. The blue line represents the sinusoidal fit using the estimated period. The middle panel shows a visualisation of the distribution of the periods. The highest peak represents the highest number of periods occurring in the simulations. The x-axis is divided in a 0.1 h bin. The right one shows the periodogram analysis results of the Monte Carlo Simulated light curves. Periodograms of the 10000 synthetic light curves are shown in grey, originating in the Monte Carlo simulation of the original light curve. The black line represents the periodogram of the observed light curve. Red, blue, and green lines represent the one, two, and three $\sigma$ limits of the mean light curve in yellow. The cyan-blue vertical line corresponds to the period found in the LS periodogram of the observed light curve. The dates of the observations are mentioned in the sub caption. The mean, 1$\sigma$, 2$\sigma$ and 3$\sigma$ lines in the first figures in each panel are truncated and limited to show only the data near the resulting period.}	
 
   \label{supCT2}
\end{figure}

\begin{figure*}
\begin{minipage}[b]{.7\linewidth}

  \centering
    \includegraphics[width=.9\linewidth]{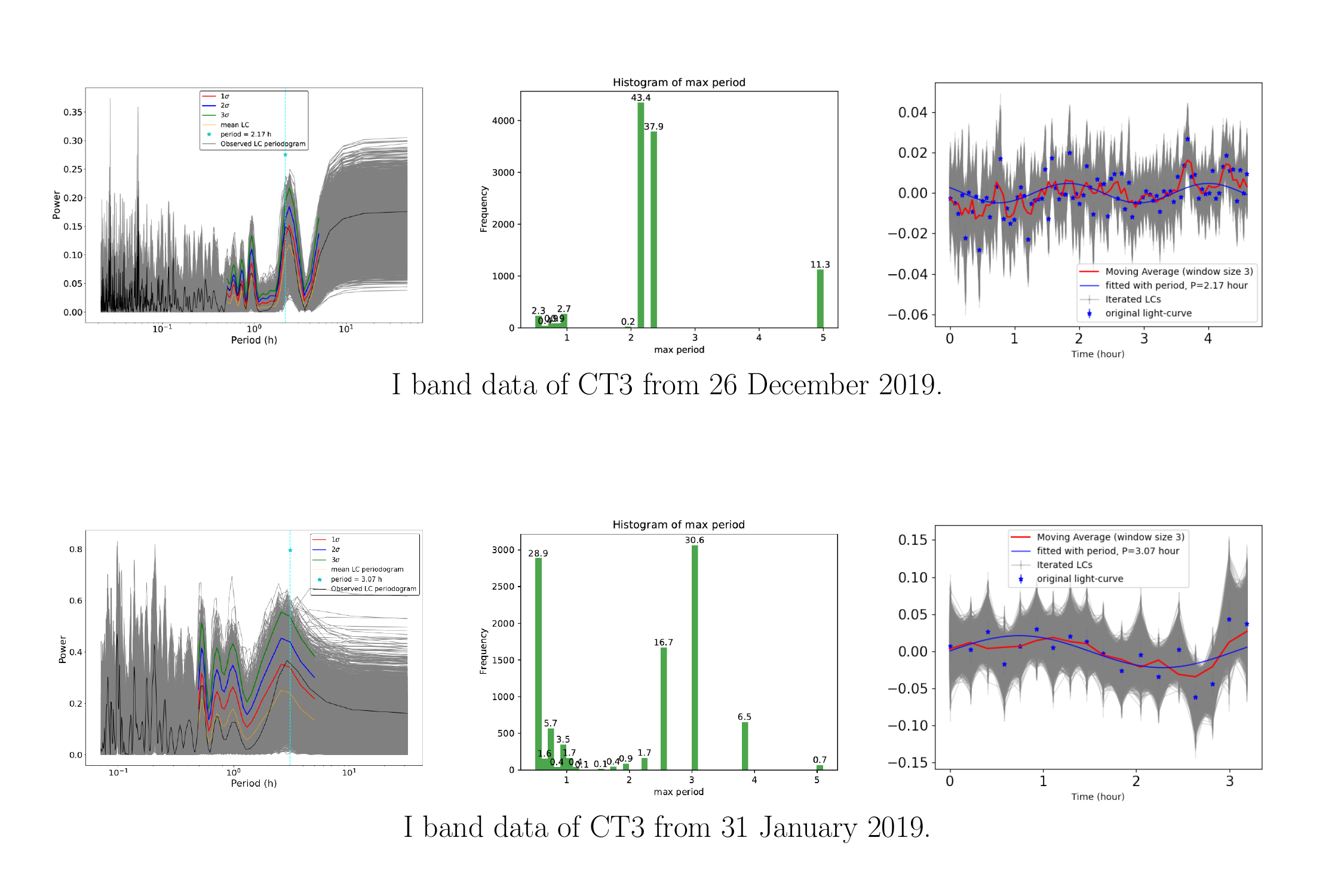}

\end{minipage}
 \caption{CT3 data are shown here. Figure descriptions, styles and colour coding is same as the previous figure.}	
   \label{supCT3}
\end{figure*}
\begin{figure*}
\begin{minipage}[b]{.7\linewidth}

  \centering
    \includegraphics[width=.9\linewidth]{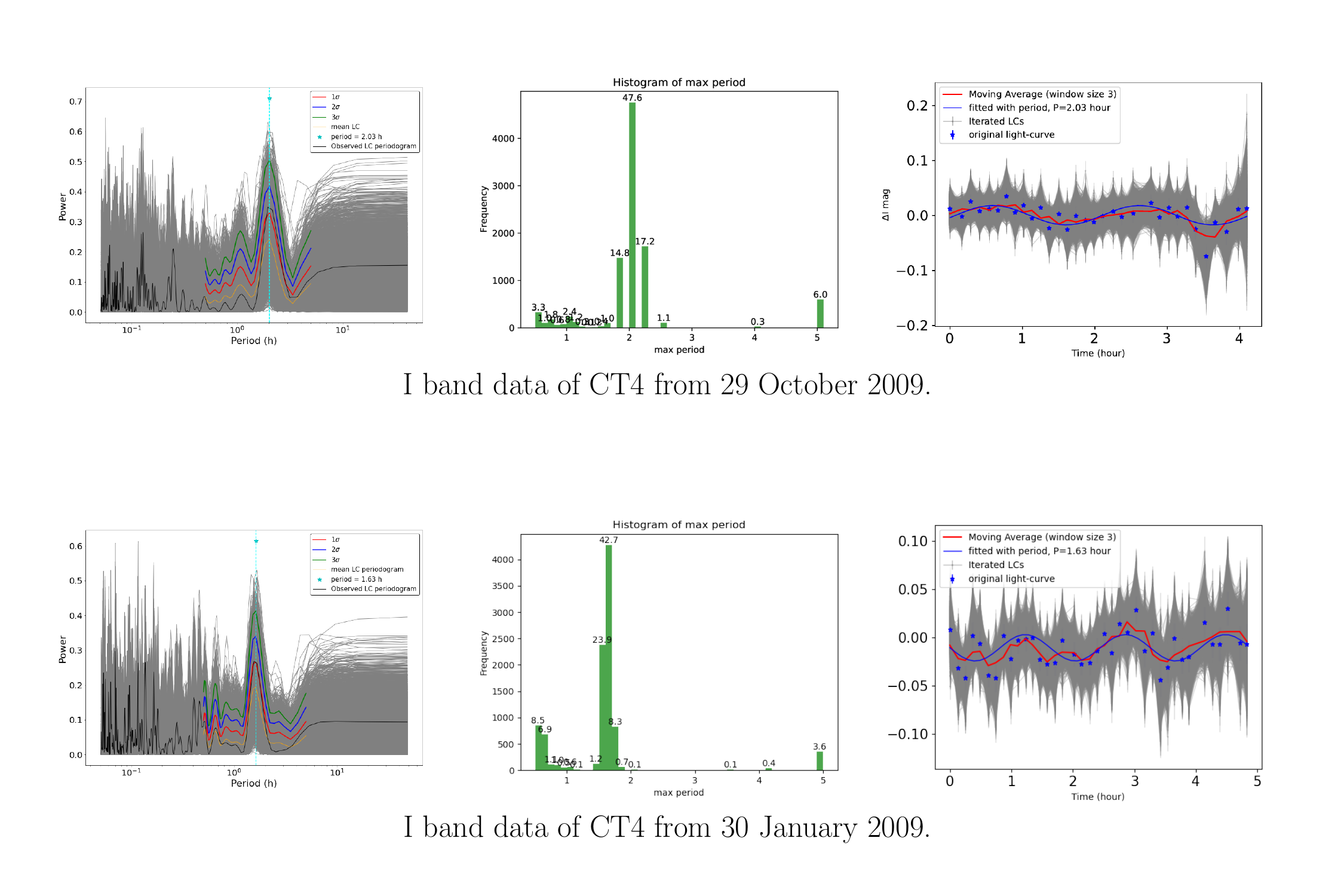}

\end{minipage}
 \caption{CT4 data are shown here. Figure  descriptions, styles and colour coding is same as the previous figure.}	
   \label{supCT4}
\end{figure*}


\end{document}